\documentclass{aa}  
\usepackage{graphicx}
\usepackage{subcaption}
\usepackage{xcolor}
\usepackage[symbol]{footmisc}
\usepackage{booktabs}
\usepackage{physics}
\usepackage{multirow}
\usepackage{amsmath}
\usepackage{hyperref}
\usepackage{url}
\usepackage{academicons}
\usepackage{svg}
\hypersetup{breaklinks=true,colorlinks=true,urlcolor=blue, linkcolor=blue,  citecolor=blue, bookmarksopen=true}
\usepackage{orcidlink}
\usepackage{txfonts}

\begin{document} 

   \title{UV-processing of icy pebbles in the outer parts of VSI-turbulent disks}

    \author{Lizxandra Flores-Rivera\inst{1}\orcidlink{0000-0001-8906-1528}
          \and
          Michiel Lambrechts\inst{1}\orcidlink{0000-0001-9321-5198}
          \and
          Sacha Gavino\inst{2}\orcidlink{0000-0001-5782-915X}
          \and
          Sebastian Lorek\inst{1}\orcidlink{0000-0002-5572-4036}
          \and
          Mario Flock\inst{3}\orcidlink{0000-0002-9298-3029}
          \and
          Anders Johansen\inst{1}\orcidlink{0000-0002-5893-6165}
          \and
          Andrea Mignone\inst{4}\orcidlink{0000-0002-8352-6635}
          }

   \institute{Centre for Star and Planet Formation, Globe Institute, University of  Copenhagen, \O ster Voldgade 5-7, 1350 Copenhagen, Denmark
   \email{lizxandra.rivera@sund.ku.dk}
    \and
    Niels Bohr Institute, University of Copenhagen, Jagtvej 155A, 2200 Copenhagen N., Denmark
    \and
    Max Planck Institut f\"ur Astronomie, K\"onigstuhl 17, 69117 Heidelberg, Germany
    \and
    Dipartimento di Fisica, Università degli Studi di Torino , Via Pietro Giuria 1, I-10125 Torino, Italy}

   \date{Submitted on November 8. Accepted.}

 
  \abstract
  {Icy dust particles emerge in star-forming clouds and are subsequently incorporated in protoplanetary disks, where they coagulate into larger pebbles up to mm in size. In the disk midplane, ices are shielded from UV radiation, but moderate levels of disk turbulence can lift small particles to the disk surface, where they can be altered, or destroyed. Nevertheless, studies of comets and meteorites generally find that ices at least partly retained their interstellar medium (ISM) composition before being accreted onto these minor bodies. Here we model this process through hydrodynamical simulations with VSI-driven turbulence in the outer protoplanetary disk. We use the \textsc{PLUTO} code in a 2.5 D global accretion setup and include Lagrangian dust particles of 0.1 and  1 mm sizes. In a post-processing step, we use the \textsc{RADMC3D} code to generate the local UV radiation field to assess the level of ice processing of pebbles. We find that a small fraction ($\sim$17$\%$) of 100 $\mu$m size particles are frequently lifted up to $Z/R=0.2$ which can result in the loss of their pristine composition as their residence time in this layer allows for effective CO and water photodissociation. The larger 1 mm size particles remain UV-shielded in the disk midplane throughout the dynamical evolution of the disk. Our results indicate that the assembly of icy bodies via the accretion of drifting mm-size icy pebbles can explain the presence of pristine ice from the ISM, even in VSI-turbulent disks. Nevertheless, particles $\leq$ 100~$\mu$m experience efficient UV processing and may mix with unaltered icy pebbles, resulting in a less ISM-like composition in the midplane.}


   \maketitle
%

\section{Introduction}

The outer disk is a region of interest as multiple chemical species have been detected at different layers in the disk \citep{oberg_2021}. Icy volatiles transported on dust grains in protoplanetary disks \citep{Henning_2013, Krijt_2023} constitute a significant portion of the ice budget in protostellar cores, accounting for approximately 26 to 60$\%$ of the total elemental oxygen along with H$_{2}$O and CO$_{2}$, whereas 14 to 27$\%$ for elemental C \citep{Boogert_2015}. The \textit{James Webb} Space Telescope (JWST) has provided a detailed and spatially resolved ice inventory, helping to trace the changes in ice composition over time and through different environmental conditions. A recent study by \citet{Sturm_2023} shows that the prominent absorption features in CO$_{2}$ and CO ices are shifted to pure ice absorption features, indicative that ices are exposed to energetic processing at low temperatures. 

Ices formed in the ISM carry a heavy isotope imprint in H, C, O, and N, compared to the solar/ISM values that made up the solar gas disk and are crucial for life as we know it \citep[see review by][]{Broadley_2022}. This signature is seen in comets sampling pristine outer-disk ices and primitive meteorites. For example, the high D/H ratios in deuterated H$_{2}$O and deuterated H$_{2}$S (hydrogen sulfide) in comet 67P/Churyumov-Gerasimenko suggest a prestellar origin for some of its ices \citep{Altwegg_2017}. Other more complex CHO-, N-, and S-bearing molecules in comet 67P/Churyumov-Gerasimenko show correlated abundances with protostellar material \citep[i.e., IRAS 16293-2422 B;][]{Drozdovskaya_2019}. This suggests that volatiles in Solar-like systems may be preserved from the protostellar phase into planetesimals \citep{Drozdovskaya_2019}. Primitive meteorites that formed closer to the Sun are subjected to more processing than comets and are estimated to contain $5-10\%$ pristine ISM material \citep{Alexander_2017} while comets contain about $10-30\%$ \citep{Mumma_2011}. This evidence suggests that significant amounts of ISM-like ice survived disk formation and evolution, becoming part of comets and, to a lesser extent, asteroids.

The formation and destruction reaction pathways of chemical species are driven by stellar radiation and thermal processes that further modify the ISM-like original composition of the disk. High energetic photons such as Ultra-violet (UV) photons from stellar radiation can induce photodissociation reactions on ice (AB$_\mathrm{ice}$ + $h\nu$ $\rightarrow$ A$_\mathrm{ice}$ + B$_\mathrm{ice}$) and photodesorption reactions (AB$_\mathrm{ice}$ + $h\nu$ $\rightarrow$ AB$_\mathrm{gas}$) to trigger the formation of simple and complex organic molecules, which are incorporated into pebbles, planetesimals and later planets. Volatiles and complex organ molecules are carried in particles that experience radial drift and vertical stirring due to disk instabilities, which ultimately influence the composition of planets.

Numerous previous models have examined the reprocessing of prestellar ices and have found that substantial reservoirs of prestellar ices are incorporated into the outer disk through infall accretion from the envelope \citep[i.e.,][]{Visser_2011, Yang_2013, Hincelin_2013, Drozdovskaya_2014, Yoneda_2016}. More recently, \citet{Bergner_2021} used a constant $\alpha$ value to represent gas turbulence within the disk and made a model where individual pebbles were covered by pristine ice mantles. They found that ice destruction is typically ineffective, except for small grains ($<100~\mu$m). This statement supports the scenario that icy pebbles ($>100~\mu$m) originate in the outer disk and subsequently migrate inwards, contributing to the formation of cometary bodies. 

Turbulence in protoplanetary disks is crucial regarding angular momentum transport and dust dynamics, which can affect planet formation. Beyond 0.1 au, where the magnetorotational instability \citep[MRI;][]{Balbus_1991} is suppressed due to low ionization rates \citep{Turner_2014, Dzyurkevich_2013}, turbulence can arise from hydrodynamic instabilities like the vertical shear instability \citep[VSI;][]{Nelson_2013}. The VSI is driven by vertical differential rotation in baroclinic disks \citep{Stoll_2017, Barker_2015} and the turbulence intensity increases with the aspect ratio of the disk in rapid cooling timescales \citep{manger_2020,manger_2021}. It causes anisotropic turbulence with $Z-\phi$ stress-to-pressure ratio typically reaching magnitudes of $10^{-2}$, while the $R-\phi$ stress-to-pressure ratio range from $10^{-4}$ to $10^{-3}$ \citep{Flock_2017, Flock_2020}, and large-scale gas circulations, which can stir up dust particles and create vortices and zonal flows \citep{Lyra_2019}. 

Global simulations that have incorporated Lagrangian particles of millimeter-sized dust demonstrated that these particle sizes can be lifted close to the gas scale height above the midplane due to strong vertical turbulence \citep{Stoll_2017, Flock_2017, Flock_2020, Dullemond_2022}. High-resolution simulations indicate that strong shear between neighboring bands of these meridional circulations can generate vortices in the $R-Z$ plane \citep{flores-rivera_2020, Fuksman_2024a, Fuksman_2024b} as well as in the $Z-\phi$ plane \citep{manger_2020, manger_2021} that are related to the Kelvin-Helmholtz instability and could serve as dust trap locations, enhancing local dust concentrations. The corrugation modes of the VSI can be translated to rings with azimuthal structures in the disk that can potentially explain dust substructures observed by ALMA \citep{Blanco_2021}.

\begin{figure} [htp]
\centering
\includegraphics[width=9cm]{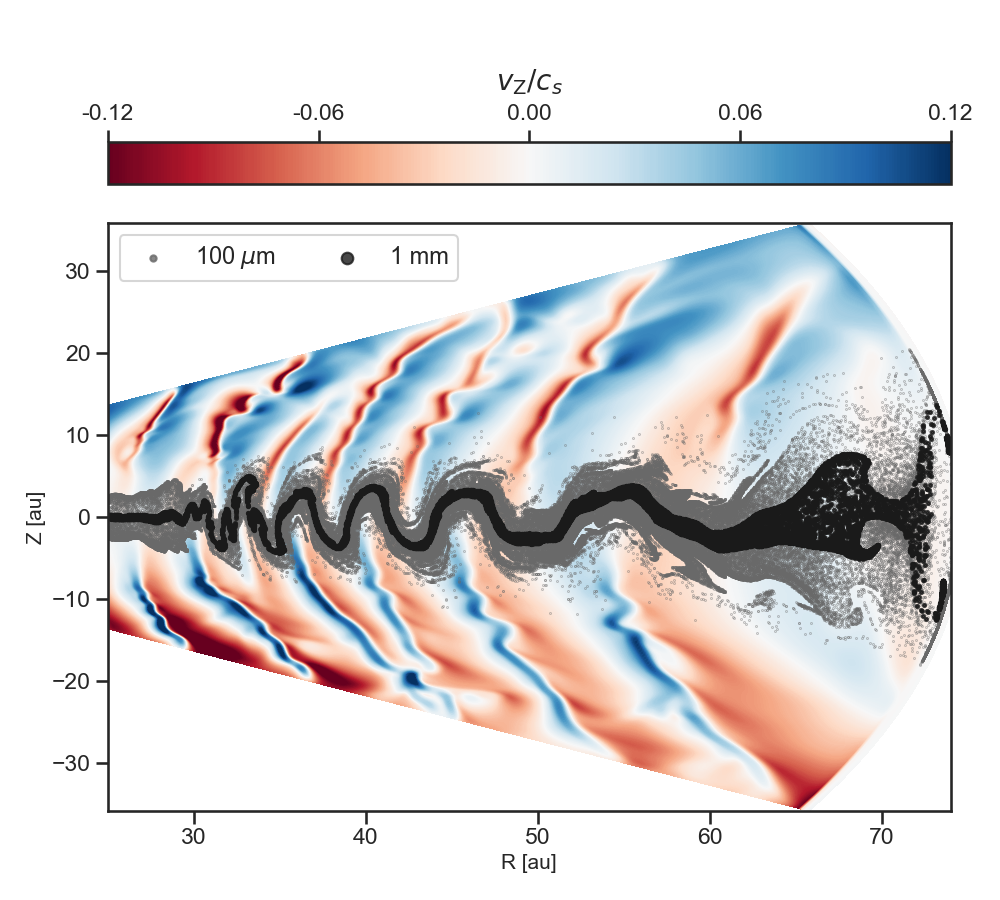}
\caption{\textit{Left} Dust particles of 100 $\mu$m (grey) being stirred by disk turbulence high above the disk midplane, compared to the more settled 1 mm particles (black). The disk turbulence is driven by the VSI represented by the vertical gas velocities ($v_\mathrm{z}$) expressed in terms of the local sound speed. This simulation snapshot at 300 orbits represents a disk time evolution for 0.1 Myrs at 50 AU. }\label{fig:vsi+dust}
\end{figure}

The radial and vertical transport by the VSI within the disk is expected to impact the dust particle trajectory significantly and consequently its ice composition. For this reason, this work is dedicated to investigating the gas and dust dynamics and providing predictions on the ice chemistry timescales in the outer disk region. Turbulence, can lift and mix the dust particles at different vertical layers, exposing them to vastly different thermal and UV environments \citep{Semenov_2011}.

The outline of the paper is described as follows. In \S 2 we describe the numerical methods for modeling accretion disk that leads to the generation of the VSI modes. In \S 3 we describe the physical properties used to \textsc{RADMC3D}. In \S 4 we describe the thermal and photochemical reactions rates used to calculate the CO ice destruction timescale. \S 5 and 6 we present the results and discussion, respectively. Finally, \S 7 we present our conclusions. 

\section{Numerical methods}
\label{sec:gas}

We use the PLUTO code\footnote{\textsc{PLUTO 4.3} is an open source code available for download: \url{http://plutocode.ph.unito.it/download.html}} \citep{Mignone_2007} to conduct the simulations of the gas in the protoplanetary disk. Our HD simulations are described based on the conservation of mass and momentum  

\begin{equation}
    \frac{\partial\rho}{\partial t} + \nabla \cdot (\rho\textbf{v}) = 0 
,\end{equation}
\begin{equation}    
    \frac{\partial(\rho\textbf{v})}{\partial t} +  \nabla \cdot (\rho \textbf{vv}) + \nabla P  =  - \rho \nabla\Phi
,\end{equation}where $\rho$ is the mass density, $\textbf{v}$ is the velocity vector, $\rho\textbf{v}$ is the momentum density vector, and $P$ is the gas pressure. We select the isothermal equation of state with $P=c_{\mathrm{s}}^{2}\rho$, where $c_{\mathrm{s}}$ is the isothermal sound speed. The simulations are in 2.5D (2 dimensions, 3 components) using spherical coordinates ($r,\theta,\phi$) with axisymmetry in the azimuth. The grid cells are set up with a logarithmic increase in the radial domain and a uniform spacing in the meridional domain. The radial domain ranges from 20 to 75 au and the vertical domain covers approximately $\pm$5 gas scale height. 

The initial gas density profile in cylindrical coordinates ($R,Z$) derived by \citet{Nelson_2013} as 

\begin{equation}
    \label{density}
      \rho(R,Z) = \rho_{\mathrm{0}} \left(\frac{R}{R_{0}} \right)^{p}  \exp\left [ \left (\frac{GM_{*}}{c_{\mathrm{s}}^{2}} \left[\frac{1}{\sqrt{R^2 + Z^2} } - \frac{1}{R} \right] \right) \right] \,, 
\end{equation}

\begin{equation}
      \label{azimuthalvelocity}
      \Omega(R,Z) = \Omega_\mathrm{k} \left[ (p+q) \left(\frac{H_\mathrm{R}}{R} \right)^{2} + (1+q) - \frac{qR}{\sqrt{R^2 + Z^2} } \right]^{1/2}  ,
\end{equation}

\noindent where $G$ is the gravitational constant, $M_{*}$ is the mass of the star, \textit{p} and \textit{q} are the radial profile exponent of the temperature and density, respectively. The initial density at the midplane is defined as, $\rho_\mathrm{0} = \frac{\Sigma_{\mathrm{0}}}{\sqrt{2\pi}~H_\mathrm{0}R_{0}}$ where $\Sigma_{\mathrm{0}}$ is the initial surface density at $R_\mathrm{0}$ is the reference radius, $H_\mathrm{0}$ is the reference scale height (see Table \ref{tab:resolution}). The Keplerian frequency is $\Omega_\mathrm{k} = \sqrt{GM_{*}/R^{3}}$. The gas scale height of the gas is defined as

 \begin{figure} [htp!]
\centering
\includegraphics[width=8.3cm]{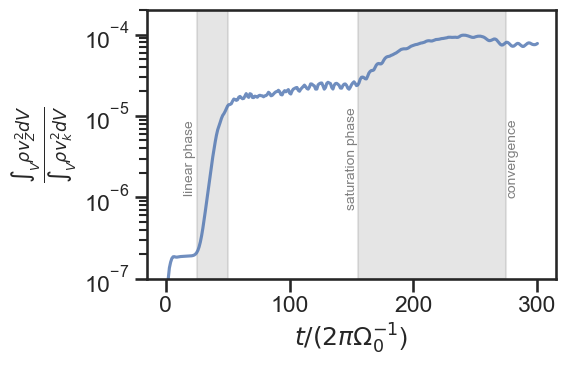}
\caption{Time evolution of the global kinetic energy of the gas at 50 au normalized with the kinetic energy from pure Keplerian velocity.}
\label{fig:vsi+dust2}
\end{figure}

\begin{equation}
      \label{eq:hr}
      H_\mathrm{R} = H_\mathrm{0}  \left(\frac{R}{R_\mathrm{0}} \right)^{(q+3)/2}   
      \,
,\end{equation}

\noindent The reference scale height ratio is $H_\mathrm{0}$ = 0.1 at $R_{0}$ (Table \ref{tab:resolution}). The temperature profile is $T\propto c_\mathrm{s}^{2}$ is the sound speed and is given based on equation \ref{eq:hr} so that $c_\mathrm{s}$=$H_\mathrm{R}$$\Omega_\mathrm{k}$. The scale length of the vertical modes is proportional to $H_\mathrm{0}$.
The vertical component of the velocity in cylindrical coordinates, $v_{Z} = v_{r}\cos(\theta) - v_{\theta}\sin(\theta)$, is used for the analysis of the VSI (Figure \ref{fig:vsi+dust}). 

\begin{table}[htb]
   \caption{Grid setup and initial parameters for the different runs.} 
   \label{tab:resolution}
   \small 
   \centering 
   \begin{tabular}{lccr} 
   \noalign{\smallskip} \hline \hline
   \noalign{\smallskip} \textbf{$N_\mathrm{r} \times N_\mathrm{\theta} \times N_\mathrm{\phi}$} & \textbf{$\Delta r$} & \textbf{$\Delta \theta$} \\  
   \hline 
   512 $\times$ 448 $\times$ 1  & 0.20:0.75 & 1.0707:2.0707 \\  
   \hline
\multicolumn{3}{c}{Initial parameters} \\
   \hline

\textit{p} & -1.5 \\ 
\textit{q} & -1.0 \\
\textit{$R_\mathrm{0}$}[au] & 50.0 \\
\textit{$H_\mathrm{0}$} & 0.1 \\
$\Sigma_\mathrm{0}$ [g~cm$^{-2}$] & 100\\
\noalign{\smallskip} \hline \noalign{\smallskip}

\end{tabular}

\end{table}

\subsection{Dust}
\label{sec:init_cond}

\begin{figure*}
\includegraphics[width=18cm]{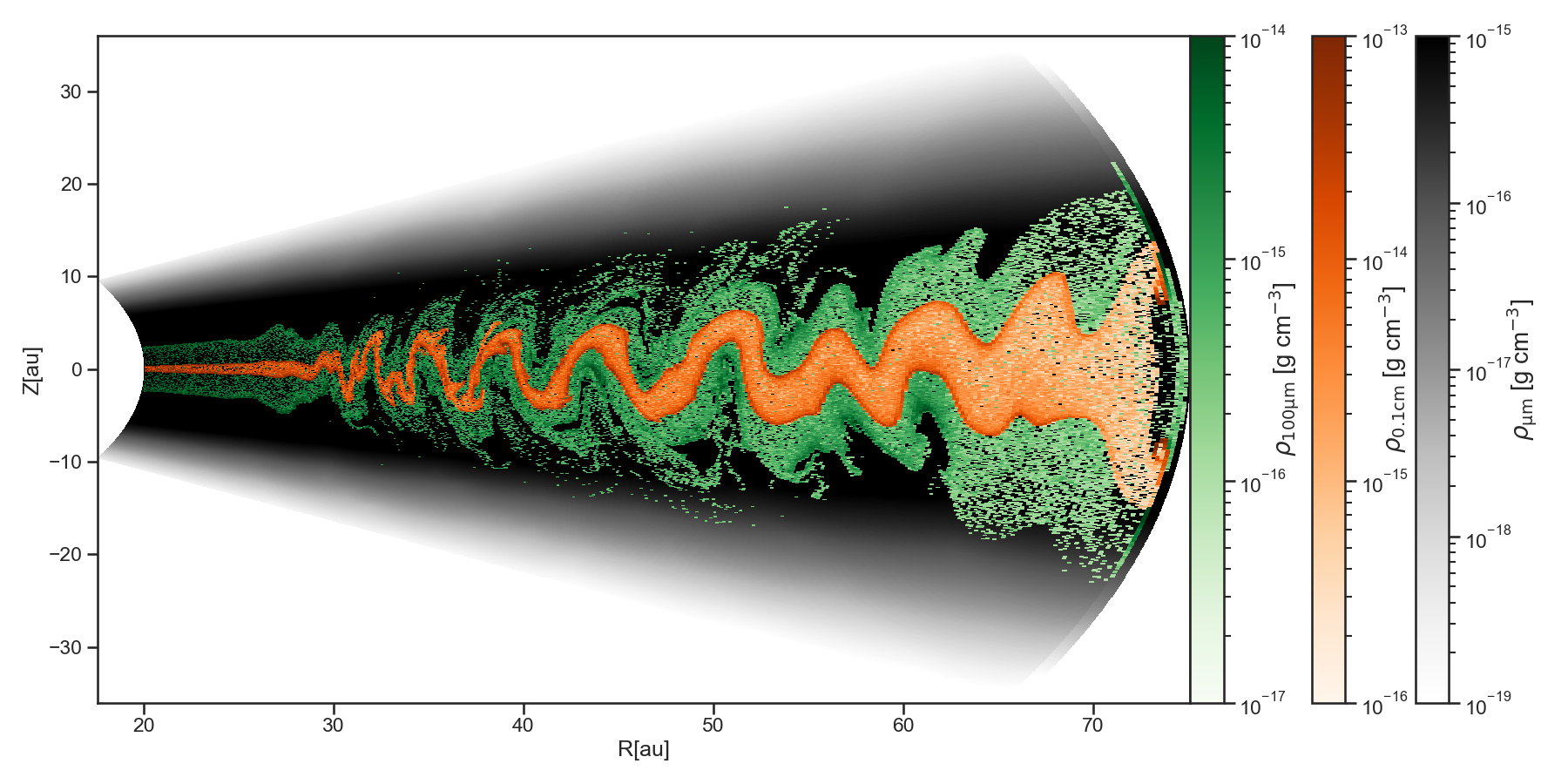}
\caption{Dust density distribution of our model at 200 orbits. The large grains are 100 $\mu$m (green) and  1 mm (orange). The small grains spanning from 0.1-30 $\mu$m is plotted as a black shade in the disk and are assumed to be well-coupled with the gas.}
\label{fig:dust_dens}
\end{figure*}

In this paper, we are interested in how the dust particles are stirred up by the VSI turbulent motions (Figure \ref{fig:vsi+dust}) and whether they are processed in different vertical layers, closer to high UV flux regions in the outer disk. From the 2D axisymmetric model of the density and rotational velocity described in \S\ref{sec:gas},  we introduced two Lagrangian particles with sizes of 100 $\mu$m and 1 mm in the simulation, and their vertical dust distribution is described as a Gaussian profile. Initially, the particles are distributed within a scale height close to the gas scale height to capture the initial dust-settling phase before grains settle more in the midplane. The particle dynamics is described using a Lagrangian dust module \citep{flock_2021} in the PLUTO code using a spherical coordinates system. These dust grains are assumed to be homogeneous, compact, and spherical with a dust density of, $\rho_\mathrm{dust} = $1.7 g~cm$^{-3}$. We do not include the effects of the dust-to-gas feedback. We also assume that dust particles are small enough compared to the gas mean free path length so they can be treated in the Epstein regime \citep{Epstein_1924}. The stopping time is important as the interaction that dust particles have with gas is different depending on the dust size. The stopping time dictates the time that it takes for a dust particle to be slowed down by the gas flow and it is defined as

\begin{equation}
    \label{stopping_time}
      t_{s} = \frac{\mathrm{St}}{\Omega_\mathrm{k}} = \frac{a\rho_\mathrm{dust}}{\rho(R,Z) H_\mathrm{R} \Omega_\mathrm{k}} = \frac{a\rho_\mathrm{dust}}{\rho(R,Z) c_{s}}\, 
\end{equation}

\noindent where $a$ is the dust particle size, which is kept constant throughout the dynamical evolution of the disk. The Stokes number can be rewritten in terms of the disk physical parameters,

\begin{equation}
    \label{stokes}
      \mathrm{St} = \frac{a\rho_\mathrm{dust} \sqrt{2\pi}}{\Sigma_{0}} \exp \left [ \frac{Z^{2}}{2H_\mathrm{R}^{2}} \right ] \left(\frac{R}{R_{0}}\right)^{1/2} \, 
\end{equation}

The dynamical trajectory of our particles will be influenced by their sizes and the turbulent gas motion. For the midplane, where $Z/H_\mathrm{R}$=0, $\exp[...]$=1, and at $Z/H_\mathrm{R}$=1, $\exp[...]$$\sim$1.65, therefore, for one scale height variation the Stokes number or particle size changes by less than a factor of 2. For example, the Stokes number for a 0.1 cm (1 mm) particle size at the midplane at 50 au is St=0.00426, and at $Z=5$ au, the St=0.0070. Similarly, the Stokes number is an order of magnitude lower for the smaller size of 0.01 cm ($100~\mu$m). This is further motivated if we consider the turbulent fragmentation velocity assumed to be 1 m~s$^{-1}$ and $\alpha=10^{-3}$ (a typical level of turbulence consistent with VSI), for which the fragmentation-limited Stokes number results in 0.0019 with a corresponding particle size of 0.045 cm at the midplane \citep[see equation 36 in][]{Birnstiel_2023}. This implies that particles are small enough to avoid destructive collisions, which enables growth to larger sizes through mechanisms such as sticking and coagulation sticking collisions until the fragmentation barrier is reached. 
We inject a total of $10^{6}$ Lagrangian particles in the simulation uniformly distributed over two large grain sizes, and uniformly distributed in the $r-\theta$ plane. The initial velocity of the particles is adjusted to align the local Keplerian velocity.

\subsection{Dust mass distribution of large and small dust particles}

We follow the approach in \citet{Ruge_2016} to calculate the mass distribution between the small and large dust grains as seen in Figure \ref{fig:dust_dens}. For this, we first use the dust-size distribution \citep{Mathis_1977}

\begin{equation}
    \label{dust_size_dist}
      \frac{dn(a)}{da} \propto a^{-3.5} \, 
\end{equation}

We do not model the small grains as Lagrangian particles in the VSI simulations, as these will remain well-coupled to the gas. We use grain sizes between 0.1 $< a_\mathrm{small} < 30$ $\mu$m discretized in 10 bins. The number 10 is arbitrary as it is only used to determine the dust-to-gas mass fraction in small size bins. These dust particles are well-coupled with the gas motion and are distributed uniformly in the radial and meridional domain. The mass of small dust, $M_\mathrm{small,d}$, can be determined

\begin{equation}
    \label{small_dust_mass}
      M_\mathrm{small,d} = \frac{4}{3} \pi \rho_\mathrm{dust} \sum_{i=1}^{10} a_\mathrm{small,\textit{i}}^{3} N_i(a_{i}) \, 
\end{equation}

\noindent where $N(a_i)$ is the absolute number of particles with radius $a_i$. The $N(a_i)$ can be expressed as a proportion relative to a reference number of particles. We select the total amount of particles, $N_\mathrm{max}$ with radius $a_\mathrm{max}$:

\begin{equation}
    \label{dust_size_dist_rewritten_small}
      N_i(a_i) = N_\mathrm{max} \left( \frac{a_i}{a_\mathrm{max}} \right)^{-2.5}\, 
\end{equation}

Similarly the amount of mass for the large particles is described

\begin{equation}
    \label{large_dust_mass}
      M_\mathrm{large,d} = \frac{4}{3} \pi \rho_\mathrm{dust} \sum_{g=1}^{2} a_\mathrm{large,\textit{g}}^{3} N_g(a_{g}) \, 
\end{equation}

where

\begin{equation}
    \label{dust_size_dist_rewritten_large}
      N_{g}(a_g) = N_\mathrm{max} \left( \frac{a_g}{a_\mathrm{max}} \right)^{-2.5}\, 
\end{equation}

The total mass of small and large grains is

\begin{equation}
    \label{Mtot}
      M_\mathrm{large,d} + M_\mathrm{small,d} = \frac{4}{3} \pi \rho_\mathrm{dust} \frac{N_\mathrm{max}}{a_\mathrm{max}} \left( \sum_{i=1}^{10} a_{i}^{0.5} + \sum_{g=1}^{2} a_{g}^{0.5}  \right) \, 
\end{equation}

In the case of large dust grains, each individual "test" particle contains a certain number of dust particles, $\tilde{N}_{g}(a_g)$. Finally, the swarm particle mass for each large particle is given by

\begin{equation}
    \label{ng}
      \tilde{N}_{g}(a_g) = \frac{N_{g}(a_g)}{10^{6}} \, 
\end{equation}

\noindent where $10^{6}$ is the total number of large particles. For a given distribution we obtain the dust-to-gas mass ratio of $M_\mathrm{small,d}/M_\mathrm{gas}=0.0003$, $M_\mathrm{100 \mu m,d}/M_\mathrm{gas}=0.0015$, and $M_\mathrm{1mm,d}/M_\mathrm{gas}=0.0036$. 

\section{Radiative-transfer as post-processing}
\label{sec:radmc3d}


We use hydrodynamical simulations to obtain the evolution of the particle component due to settling, drift, and stirring caused by VSI turbulence. For simplicity, these simulations use a local isothermal temperature profile. Using RADMC-3D \citep{Dullemond_2012}, we compute the local UV field and the dust temperature for each dust species to calculate ice processing on dust grains as a post-process step. We do not attempt to fit our physical modeling structure with actual observations. We rather use the stellar template as a model for the stellar properties required by RADMC3D to reproduce the dust temperature and radiation field profiles. 

Our fiducial case is described using the TW Hydra stellar template, a well-known low-luminosity ($L_{*}$ = 0.2 L$_{\odot}$ and $L_\mathrm{UV} \sim 10^{-3}$ L$_{\odot}$) T Tauri star with a $M_{*} = 0.8 ~M_{\odot}$, $T_\mathrm{eff} = 4110$ K, and $R_{*} = 1.04 ~R_{\odot}$ \citep{Andrews_2012}. We also attempted to do a "test" comparison with another source with a higher UV field to explore the photodesorption and photodissociation rates of particles subjected to ice processing. Note that by changing the stellar properties one ought to adjust the disk physical properties accordingly; however, performing a parameter study of the disk properties in conjunction with the stellar properties is out of the scope of this study. This second source is the high luminosity ($L_{*}$ = 40 L$_{\odot}$ and $L_\mathrm{UV} \sim 10^{-1}$ L$_{\odot}$) star, AB Aur (see Appendix \S\ref{sec:tw_hydra_spec} for more details about its stellar properties). Both sources have different UV intensity fields. The total integrated UV luminosity for the high luminosity case is two orders of magnitude greater than that of the low-luminosity case. 

\begin{figure*}[hpt!]
\centering
\includegraphics[width=18cm]{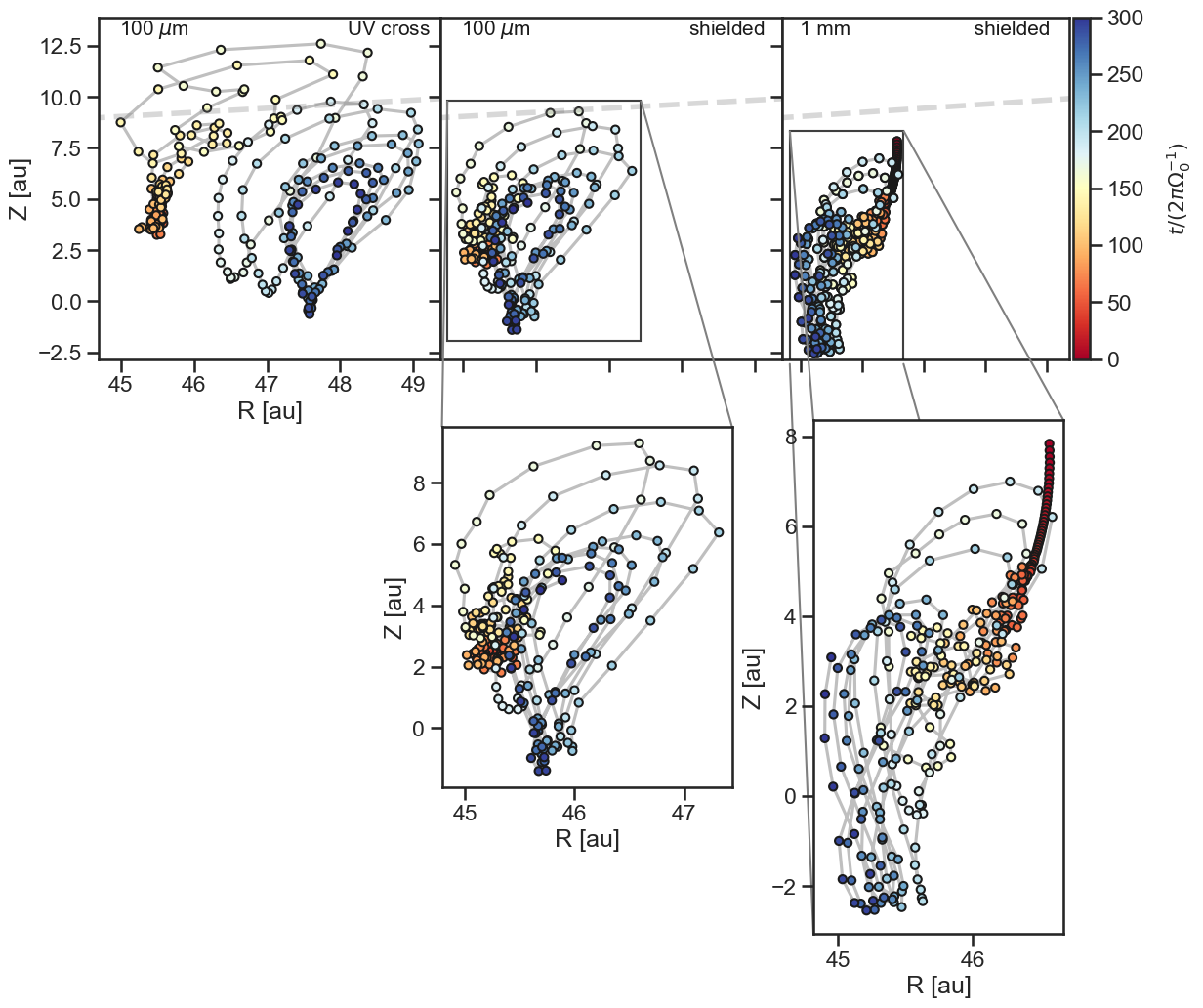}
\caption{Spatial distribution of the dynamical trajectory for three different particles. The \textit{top} panels show the track of the particles and the colorbar represents the dynamical evolution in orbits. The dashed-gray line marks the elevation of $Z/R = 0.2$, our definition of the bottom most UV layer. The \textit{top, left} and \textit{top, middle} panel shows the dynamical trajectory of two 100 $\mu$m size particle, one showing that it crosses the $Z/R = 0.2$ during the saturation phase of the VSI  after $\sim$100 orbits, and the other one being shielded all the dynamical evolution of the disk. The \textit{top, right} panel shows the trajectory for a 1 mm particle being shielded in the disk all the time. Our results show no 1 mm particle crosses $Z/R = 0.2$. At the beginning of the simulation, all the particles experience dust settling towards the midplane and inward radial drift. After $\sim$40 orbits, VSI induces diffusion in the radial and vertical motion of the particles.}
\label{fig:particle_trajectory_gallery}
\end{figure*}

In order to determine the dust opacity, our small dust grain sizes, 0.1-30 $\mu$m, are assumed to be well-coupled with the gas, sharing the same spatial distribution as the gas. The large grains of sizes 100 $\mu$m and 1 mm are, although, more settled in the disk, both particle sizes experience much stirring by the upward and downward motions of the VSI, clearly seen in Figure \ref{fig:vsi+dust} and Figure \ref{fig:dust_dens}. We use the \textsc{DIANA} standard dust opacities that consider a mixture of $60\%$ amorphous laboratory silicates, $15\%$ amorphous carbon and $25\%$ porosity by volume, well-mixed on small scales \citep{Woitke_2016}. We performed another run using the DSHARP opacities and the temperature profile is about a factor of 0.94 lower than the temperature profile with the adopted \textsc{DIANA} opacities (see more Appendix \ref{sec:opacities}). We do not include ice-coated grains in our models, that could show characteristic opacity features at $\sim$3 $\mu$m \citep[i.e.,][]{Ossenkopf_1994}. Here we focus on the analysis of the ice processing timescale compared to the dynamical timescale that the particles spend above certain vertical spatial height in the disk (see Figure \ref{fig:particle_trajectory_gallery}). 

The local radiation field is calculated at each grid point in our disk model. The fluxes are computed at 71 wavelength bins spanning 91-200 nm. Similarly as described in \citet{Bergner_2021}, the wavelengths are chosen such that, given the unattenuated stellar spectrum, the total wavelength-summed photodissociation differs by less than 5$\%$ from the wavelength-integrated photodissociation rate for the molecules considered here. We do not consider the scattering effects on the propagation of Ly-$\alpha$ photons, which has been shown to show some enhancement deeper disk layers with the Ly-$\alpha$ regime \citep{Bethell_2011}. For this reason, our treatment for the photodesorption and photodissociation rates may be underestimated here. The resulting temperature and radiation field profiles are shown later in \S \ref{sec:ice_processing}.

\section{Chemical reaction rates}
\label{sec:chemical_react_rates}

Our chemical modeling framework does not include multiple icy monolayers in the mantle on the grains. However, we assume the particles are icy grains with a CO and H$_{2}$O composition reminiscent of protostellar conditions. We focus on the chemical reaction rates in s$^{-1}$ that quantify the ice destruction timescales of dust grains through thermal- and photodesorption as well as photodissociation. The dust surface composition is given by the adopted binding energies representative of the first generation molecule \citep[see][]{Bergner_2021}.

The thermal desorption reaction rate from the ice surface is

\begin{equation}
\centering
\label{thermal_desorp}
      k_\mathrm{th,des} = \nu \exp\left[ \frac{-E_\mathrm{des}}{k_B T_\mathrm{dust}} \right] \,, 
\end{equation}

\noindent where $T_\mathrm{dust}$ is the temperature of the small dust grain and $E_\mathrm{des}$ is the binding energy for CO which is taken to be 1150 K \citep{Garrod_2013}. We exclude to map the thermal desorption for H$_{2}$O as it only becomes important inside 20 au, even for the high luminosity "test" case. The pre-exponential factor, $\nu$, is the first-order characteristic vibrational frequency of specie CO on the dust grain \citep[see][]{Hasegawa_1992} and is defined as 

\begin{equation}
\centering
\label{nu}
      \nu = \sqrt{ \frac{2 N_s k_B E_\mathrm{des}}{ \pi^{2} m_\mathrm{CO}  }}\, ,
\end{equation}

\noindent where $N_s=10^{15}$~cm$^{-2}$ is the number of binding energy sites per surface area of the grain, $k_B$ is the Boltzmann constant, and $m_\mathrm{co}=4.65\times10^{-23}$g is the mass of CO. For CO, $\nu$ is on the order of just a few $10^{12}$ s$^{-1}$.

We also calculate the photodesorption induced by UV photons: 

\begin{equation}
\centering
\label{phot_desorp}
      k_\mathrm{ph,des} = Y_\mathrm{pdes} \sigma_\mathrm{des} F_\mathrm{UV}/2   \,. 
\end{equation}

\noindent Here, $\sigma_\mathrm{des} = \frac{\pi a^{2}}{N_b} = \frac{1}{N_s} = 10^{-15}$~cm$^{2}$, is the so-called standard photodesorption geometric cross-section for the small grains \citep[see][]{Bergner_2021} and $N_b = N_s \pi a^{2} = 10^{6}$ when using $N_s=10^{15}$~cm$^{-2}$ as the number of binding sites per grain surface for a standard grain size of 0.1 $\mu$m \citep{Hollenbach_2009}. The $Y_\mathrm{pdes} = 10^{-4}$ in units of molecules per grain per photon is the photodesorption yield assumed to be constant for CO and H$_{2}$O based on laboratory experiments \citep{Visser_2011, Öberg_2009}. A constant photodesorption yield is assumed because it is difficult to assign a molecule-specific yield as photon absorption happens under the ice layer \citep[see discussion in][]{Ruaud_2016,Bertin_2012,Munoz_2010}. The factor 1/2 reflects that the UV photons reach the dust grain  from one direction as in \citet{Bergner_2021}. The $F_\mathrm{UV}$ is the total UV flux integrated from 91 nm to 200 nm. 

We include the photodissociation rate as 

\begin{equation}
\centering
\label{phot_diss}
      k_\mathrm{ph,dis} = \sum^{}_{n} \bar{\sigma}_\mathrm{pdiss,\textit{n}} \overline{F}_{\mathrm{UV},n}/2 \,, 
\end{equation}

\noindent where $\bar{\sigma}_\mathrm{pdiss,\textit{n}}$ is the average UV photodissociation cross-sections for the CO and H$_{2}$O molecule that is taken from the Leiden Observatory database\footnote{\url{https://home.strw.leidenuniv.nl/ ~ewine/photo/}} \citep{Heays_2017} and $n$ is the wavelength bins sampled in our model. The cross-sections are scaled down to by a factor of 10 to approximate the ice phase rates since photodissociation products can be trapped in the surface and rapidly react to reform the parent molecule lowering the effective photodissociation cross-section compared to gas phase \citep[i.e.,][]{oberg_2016}.

\section{Results}

\begin{figure*}[hpt!]
\centering
\includegraphics[width=17cm]{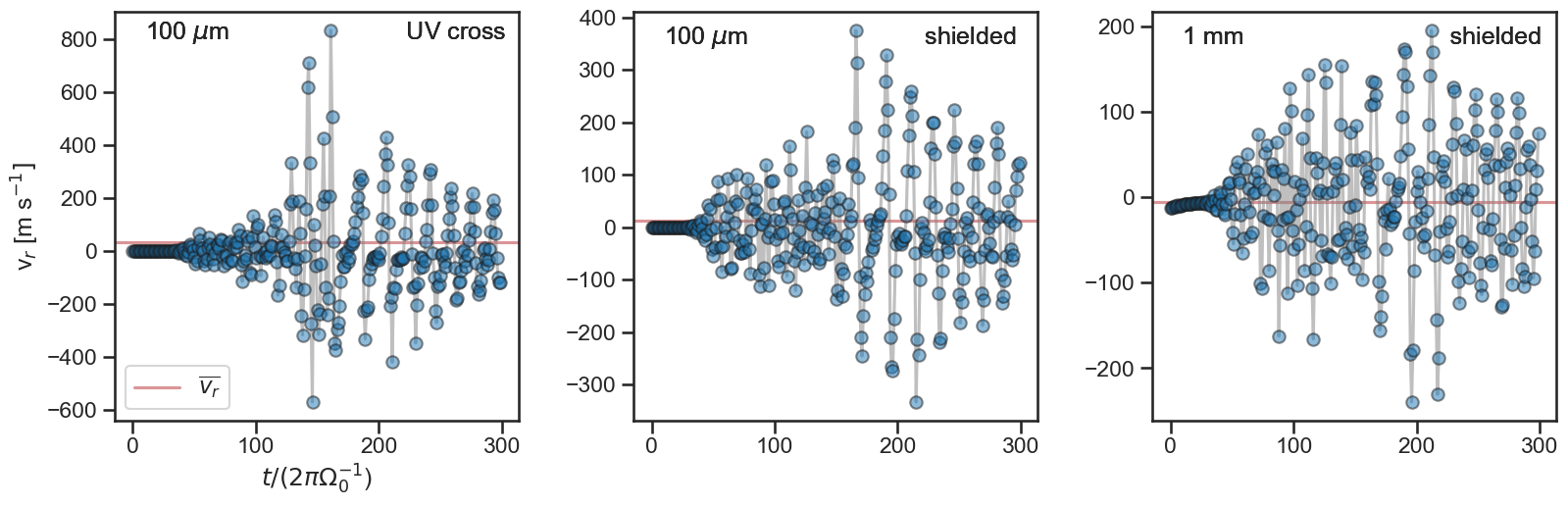}
\caption{Time evolution of the radial velocity in m~s$^{-1}$ for the same particle locations as in Figure \ref{fig:particle_trajectory_gallery}. The red horizontal line represents the mean radial velocity of the particles, 33.6 m~s$^{-1}$, 12.5 m~s$^{-1}$, and -7.7 m~s$^{-1}$, after 40 orbits, respectively.}
\label{fig:time_evolution_vr_gallery}
\end{figure*}

\subsection{Dynamical trajectory of particles}
\label{sec:dynamical_trajectory_particles}

The particles analyzed are subject to turbulence induced by the VSI. The initial distribution of the large grains is close to the gas scale height to reproduce the initial dust-settling phase for young disks. To analyze their dynamical trajectory, around 5,000 particles of both sizes were randomly selected, from a set initially localized within 40 au and 50 au radial distances from the star for all the vertical extents. We do not consider choosing particles close to the inner and outer boundaries to avoid any boundary condition effects. The trajectory of the particles reached a dynamical evolution time of 300 orbits, corresponding to a time evolution of 0.1 Myrs at 50 au. The linear phase of the VSI starts off after $t>1$ orbit from the initial radius at the inner boundary (Figure \ref{fig:vsi+dust}). Initially, all particles go under the process of dust settling towards the midplane until the VSI acts upon the particles at slightly later times, usually at the first $\sim$30 or 40 orbits, depending on the initial radial location. During the linear phase of the VSI, the particles experience small kicks or radial and vertical displacements. After $t=40$ orbits, the generation of the VSI modes continue to generate turbulence through the entire radial extension. At $t=$70-100 orbits, also called the non-linear phase, the VSI modes reach convergence and saturation in turbulence levels of $\alpha\approx10^{-3}$. The tracks of the particles at the non-linear VSI phase tell us that particles with particles size of 1 mm and 100~$\mu$m undergo significant vertical stirring when the particles are subject to turbulence in the disk (see Figure \ref{fig:particle_trajectory_gallery}). We also tried to reproduce a map of the local kinetic energy and local 1 mm dust density over time and height at 50 AU (see Figure \ref{fig:time_evol_ke_over_height}). From the top panel in Figure \ref{fig:time_evol_ke_over_height}, we depict the common features in the local kinetic energy as vertical lanes, so-called finger modes, within the first few tens of orbits corresponding to the VSI linear phase. Then, the body modes appear during the merging of the finger modes and the local kinetic energy increases slowly until the VSI reaches the saturation phase at around 150 orbits. The local dust density follows the VSI modes. The dust scale height reaches equilibrium at 5 $H_\mathrm{R}$ (red-dashed line in Figure \ref{fig:time_evol_ke_over_height} \textit{middle} panel) after 250 orbits during the saturation phase. In this stage, the width of the dust density distribution is around 2.3 $H_\mathrm{R}$.

Figure \ref{fig:particle_trajectory_gallery} shows the spatial track of three selected particle cases to illustrate different evolutions. Two $100~\mu$m particles were chosen at different vertical extents (\textit{left} and \textit{middle} panels, respectively). A 1 mm particle is also shown being shielded close to the disk midplane (\textit{right} panel). The gray dashed line corresponds to the $Z/R=0.2$ layer (also called the UV layer, see UV radiation field map in Figure \ref{fig:physical_structure} for reference), defined to be the bottom-most layer of the UV field and supported by recent observations of the CN (1-0) emission layer that is found to be a good indicator of UV penetration in the MAPS IV/ALMA imaging and fitting data from \citet{Bergner_MAPS_2021}. This $Z/R=0.2$ layer is the reference layer we use to quantify the amount of particles subject to some degree of ice processing. A particle that crosses this UV layer at any dynamical evolution time is susceptible to some degree of UV radiation. Therefore, some of its pristine composition could be susceptible to a photodestruction process (more detailed analysis in \S \ref{sec:ice_processing}). None of the 1 mm particles cross the $Z/R=0.2$ layer in our drawn particle sample. The dynamical track frame for the $100~\mu$m particle (\textit{left} panel) shows that the particle crosses the UV layer during the non-linear phase of the VSI (seen in orange color in the colorbar from Figure \ref{fig:particle_trajectory_gallery}), whereas a particle below the UV layer remains shielded from the UV radiation keeping, in principle, its pristine composition in the outer disk at least within 300 orbits. Since our analysis of the composition of the particle is limited, quantifying the number of times per orbit each particle crosses the UV layer gives a very clear understanding of the exposure time that the particles are subject to UV radiation and, therefore, more likely to photodestruction processes that affect their pristine composition (see more details in \S \ref{sec:ice_processing}).

\subsection{Quantification of particles crossing the UV layer}

From the same drawn particle sample, which includes the 1 mm and 100 $\mu$m size particles, we quantify the number of particles crossing the $Z/R=0.2$ layer. As described earlier, these particles were initially located within a radial range of $40 < R < 50$ au. However, at 300 orbits, these particles would end in different radial locations that could go beyond the initial radial range condition. At 300 orbits, we found that the percentage of particles crossing the $Z/R=0.2$ layer is about $21\%$ (1120 particles from the random selection sample), and $79\%$ of the particles are UV-shielded in the disk. We emphasize that none of the 1 mm size particles in the sample cross the $Z/R=0.2$; therefore, all particles in the \textit{UV crossing} category are 100~$\mu$m size particles. The 1 mm particles remain shielded in the disk for all the dynamical disk time considered here.

The occurrence rate at which particles cross the $Z/R=0.2$ layer happens mainly during the non-linear phase of the VSI, after 150 orbits, when the instability reaches turbulence convergence (see Figure \ref{fig:factor_time_spent_above_UVlayer} in Appendix \ref{sec:occurence_rate}). This could suggest that even at later times, beyond 300 orbits, fewer particles can still be vertically lifted at $Z/R=0.2$, but less likely as according to Figure \ref{fig:time_evol_ke_over_height} the dust scale height reaches equilibrium with the gas. From this \textit{UV crossing} classification, we cannot infer the degree of ice processing these particles are subject to. For this, we have to compare the total time that particles spent above the $Z/R=0.2$ layer, which involves the calculation of the orbital period from their radial location in the disk, together with the occurrence rate at which they are above the $Z/R=0.2$ layer (more details and analysis are described in \S \ref{sec:ice_processing}).

\begin{figure*}[htp!]
\centering 
\includegraphics[width=12cm]{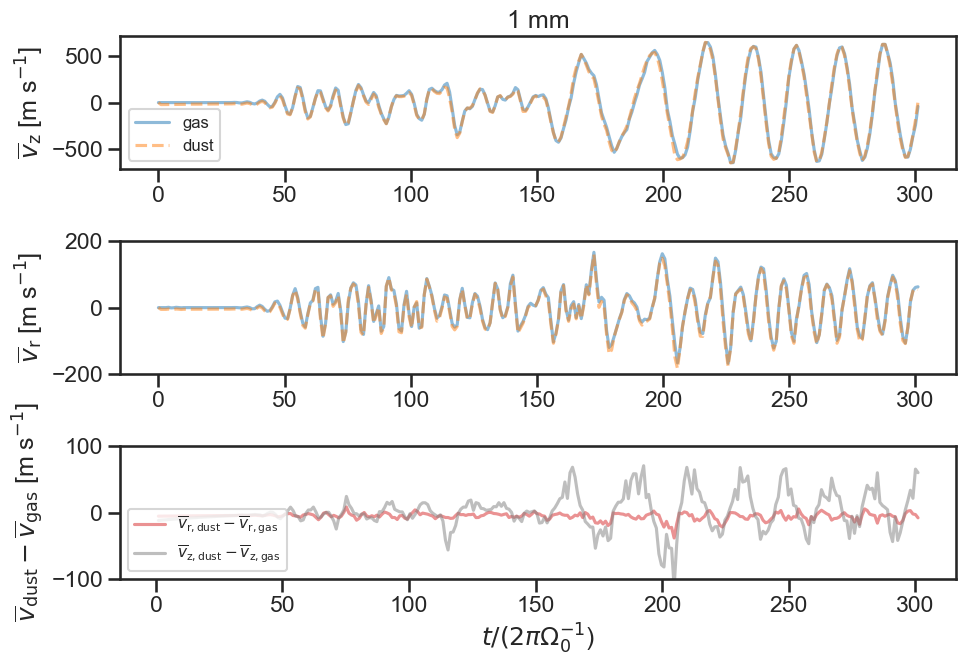}
\caption{\textit{Top and middle.} Time evolution of the mean vertical and radial velocity of a swarm of 1 mm particles around the midplane. \textit{Bottom.} The mean difference between the vertical dust and gas velocity (\textit{top} panel) in \textit{gray}, and the radial dust and gas velocity (\textit{middle} panel) in \textit{light red}.}
\label{fig:track+velo_tdrift_1mm}
\end{figure*}

\begin{figure*}[htp!]
\centering 
\includegraphics[width=12cm]{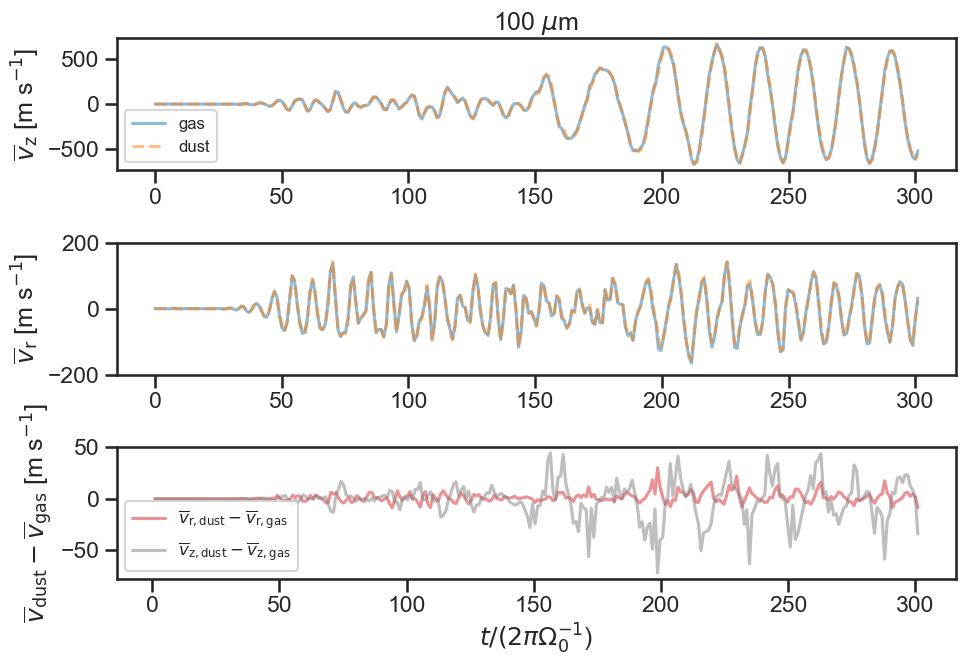}
\caption{Same as in Figure \ref{fig:track+velo_tdrift_1mm} but for a swarm of 100 $\mu$m particles.}
\label{fig:track+velo_tdrift_100microns}
\end{figure*}

\subsection{Vertical and radial velocity of particles}

To visualize and exploit the velocity profile of the particles that undergo settling, radial drift, and turbulence, Figure \ref{fig:time_evolution_vr_gallery} shows the time evolution of the radial velocity of the same particles. The initial dust-settling phase is hard to visualize in Figure \ref{fig:particle_trajectory_gallery} because of the crowding of the same particle track points being very spatially close with respect to its initial location; however, in the zoom-in box below of each panel in Figure \ref{fig:particle_trajectory_gallery} shows that the 1 mm dust particles settle inwards and something similarly happens to $100~\mu$m size particles. After $t>40$ orbits, when the VSI linearly grows in its turbulence strength, the particles experience radial and vertical diffusion over time.  During $40<t<110$ orbits, the average radial velocity of the particles are -3.9 m~s$^{-1}$ (\textit{left}), 5.0 m~s$^{-1}$ (\textit{middle}), and -7.7 m~s$^{-1}$ (\textit{right}). The negative sign in the velocity reflects the particle's inward motion, while the positive velocity reflects the particle's outward motion. 

Beyond $t>110$ orbits, the turbulence strength of the VSI is greater, injecting momentum into the particles, reflected as diffusion in the radial and vertical directions (Figure \ref{fig:time_evolution_vr_gallery}). The maximum particle diffusion is clearly seen in Figure \ref{fig:time_evolution_vr_gallery}, when the highest radial velocity, whether inwards or outwards, happens after $t=100$ orbits. The average outward radial velocity of the 1 mm particles are 164.7 m~s$^{-1}$ (\textit{left}), 91.8 m~s$^{-1}$ (\textit{middle}), and 62 m~s$^{-1}$ (\textit{right}). The average inward radial velocities are -108.5 m~s$^{-1}$ (\textit{left}), and -76.5 m~s$^{-1}$ for the $100~\mu$m (\textit{middle}), and -66.6 m~s$^{-1}$ for the 1 mm particle (\textit{right}). The highest radial velocity for the $100~\mu$m size particle (\textit{left}) occurs where the gas density is already very low, at $Z\sim12$ au disk elevation. On average, the gas radial velocity initially that carries the momentum outwards is similar to the dust radial velocity, and the difference between them is very minimal (about 0.01 m~s$^{-1}$). 

Figure \ref{fig:track+velo_tdrift_1mm} and Figure \ref{fig:track+velo_tdrift_100microns} \textit{top and middle} panels shows the vertical and radial velocity profile of both the gas (\textit{blue}) and the swarm of 1 mm dust particles (\textit{orange}) per orbit, respectively. The dust velocities are the mean velocity per orbit among multiple particles of 1 mm and 100 $\mu$m sizes. Initially, within the first 40 orbits, the particle starts settling towards the midplane. After that, the VSI starts to act upon the particles, and the radial and vertical back-and-forth motion of the particles follows closely the gas motion. The maximum radial and vertical velocity occurs after 150 orbits when the VSI reaches the saturation phase. At this phase, the particles experience vertical circling as upward and downward motions and short radial displacements that continue relatively constant until 300 orbits. The velocity profiles between the dust and gas look similar, however, Figure \ref{fig:track+velo_tdrift_1mm} and Figure \ref{fig:track+velo_tdrift_100microns} \textit{bottom} panels show the mean difference between the radial dust and gas velocity in \textit{red} and the mean difference between the vertical dust and gas velocity in \textit{gray}. The most significant difference between remains on the 1 mm particles, which is expected as these particles would have a greater de-coupling with the gas compared to the 100 $\mu$m particles. The difference in the mean vertical velocity is also greater than the mean radial velocity in both particle sizes.

\subsection{Ice particle processing by thermal and by photo-destruction processes in the outer disk}
\label{sec:ice_processing}

\begin{figure*}[hpt!]
\centering
\includegraphics[width=14cm]{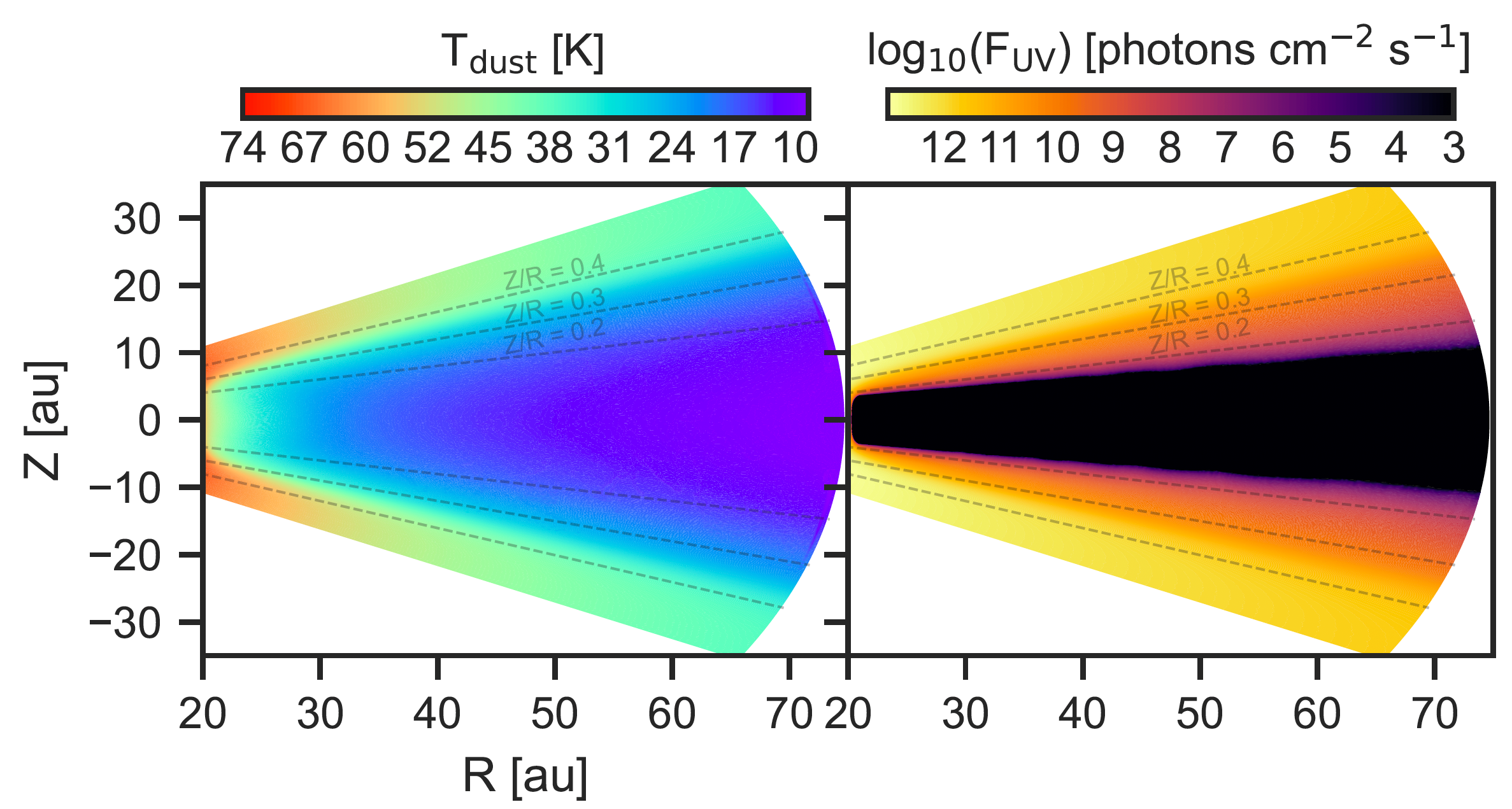} 
\caption{Dust temperature and UV radiation field for small grains of the disk properties described in \S\ref{sec:radmc3d}. The spectral shape is representative of a stellar template with a L$_{*} = 0.2~$L$_{\odot}$ and $L_\mathrm{UV} = 10^{-3}~$L$_{\odot}$. The UV flux is integrated from 91 to 200 nm.} 
\label{fig:physical_structure}
\end{figure*}

To quantify the level of ice processing in the particles, we model the local temperature profile and the UV radiation field using \textsc{RADMC3D}, as described in \S \ref{sec:radmc3d}. Figure \ref{fig:physical_structure} shows the physical structure of the low-luminosity case, $L_{*}$ = 0.2 L$_{\odot}$. The \textit{left} panel shows the temperature profile of the small grains, where grains of all sizes remain relatively cold in the midplane. Overplotted are three $Z/R$ elevations for reference. The integrated UV luminosity, from 91-200 nm, is $L_\mathrm{UV}$ = 10$^{-3}$ L$_{\odot}$ (Figure \ref{fig:physical_structure}, \textit{right} panel). The $Z/R=0.2$ level is located already in a UV-active region corresponding to log$_{10}(\mathrm{F_{UV}})\sim8.0$~photons s$^{-1}$ cm$^{-2}$, where icy particles would already be influenced by UV radiation if crossing this layer. We adopt the $Z/R=0.2$ layer to be the bottom-most UV layer. As described in \S \ref{sec:radmc3d}, we additionally explore the physical conditions of a high UV luminosity source. The UV field for the high luminosity case penetrates deeper (see Figure \ref{fig:ab_aur_uv_field} for reference), reaching higher optical depth regions closer to the disk midplane than the $L_{*}$ = 0.2 L$_{\odot}$ case. Although the UV layer at the early stages of disk formation is highly uncertain, the combination of high disk accretion and particles being less settled, variations of the UV layer for younger sources could be extended to slightly higher disk elevation than $Z/R=0.2$. 

\begin{figure*}
\centering 
\includegraphics[width=16cm]{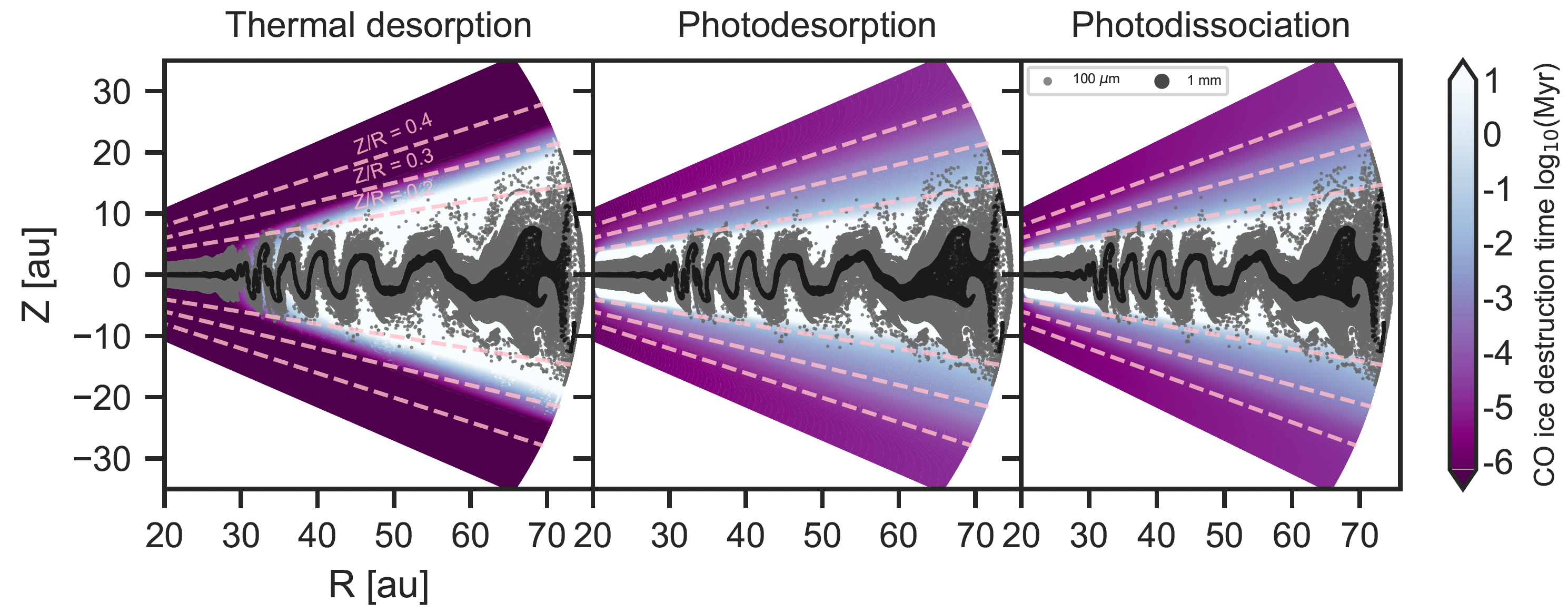} 
\caption{CO ice destruction timescale caused by thermal desorption (\textit{left},1/$k_\mathrm{th,des}$), photodesorption  (\textit{middle}, 1/$k_\mathrm{ph,des}$), and photodissociation (\textit{right}, 1/$k_\mathrm{ph,diss}$). The units are in Myrs. Overplotted is the 100 $\mu$m size particles in grey and the 1 mm size particles in black for a snapshot at 300 orbits.}
\label{fig:ice_destruction_co}
\end{figure*}

Figure \ref{fig:ice_destruction_co} (\textit{left}) shows the CO ice thermal desorption timescale for the $L_{*}$ = 0.2 L$_{\odot}$. The thermal desorption timescale, $t_\mathrm{th,des}$, is given by $t_\mathrm{th,des} = k_\mathrm{th,des}^{-1}$ (Eq.\, \ref{thermal_desorp}) for so-called first-order desorption of a mono-layer of ice \citep[i.e.,][]{Minissale_2022}.
The CO ice destruction time for the $L_{*}$ = 0.2 L$_{\odot}$ case  (Figure \ref{fig:ice_destruction_co} in \textit{left} panel) demonstrates that the CO snowline is located at around 30 au; inside this radius, the CO ice is thermally desorbed from the dust particle within a year. At $Z/R=0.2$ no particles get processed by thermal desorption. However, thermal desorption seems to be an important ice destruction reaction at elevations higher than $Z/R>0.3$ at around 50 au. For the high luminosity case, as the temperature is considerably higher, the CO snowline is pushed outwards at a radius beyond 70 au, which implies that all small and large particles get instantaneously desorbed at a more extended disk radius. For H$_{2}$O ice, thermal desorption is only effective inside 20 au, even for the high luminosity case, which is out of the radial range we consider here. Thermal desorption for CO ice and H$_{2}$O ice play a minor role in most outer disk regions, particularly at $Z/R=0.2$.

Figure \ref{fig:ice_destruction_co} (\textit{middle}) shows the photodesorption destruction timescales for CO ice in Myrs. The photodesorption timescale, $t_\mathrm{ph,des}$, is similarly set by  $t_\mathrm{ph,des}$ = $k_\mathrm{ph,des}^{-1}$ (Eq., \ref{phot_desorp}), corresponding to the first-order desorption timescale considering one active monolayer. The middle panel in Fig. 8 shows the CO photodesorption timescale in Myrs. Since photodesorption rates considered are unvarying across CO ice and H$_{2}$O ice, meaning fixed photodesorption yield and geometrical cross-section, the H$_{2}$O ice destruction timescale is identical to that of CO ice. The photodesorption time at $Z/R=0.2$ is way too long $t_\mathrm{0.2,phdes}\gg10^{4}$ years for the $L_{*}$ = 0.2 L$_{\odot}$ case, demonstrating that this photodestruction process is inefficient for the 100~$\mu$m particles that cross the $Z/R=0.2$ layer. Compared with the high luminosity case, photodesorption is very effective ($t_\mathrm{40,phdes}\sim41.7$ years), removing all ices from the dust particle at $Z/R=0.2$. Above $Z/R=0.4$, the photodesorption time is rapid ($\sim$36.5 days), and we expect all pristine CO ice and H$_{2}$O ice to be completely photo desorbed from the dust particle.

\begin{figure}[htp!]
\centering 
\includegraphics[width=8cm]{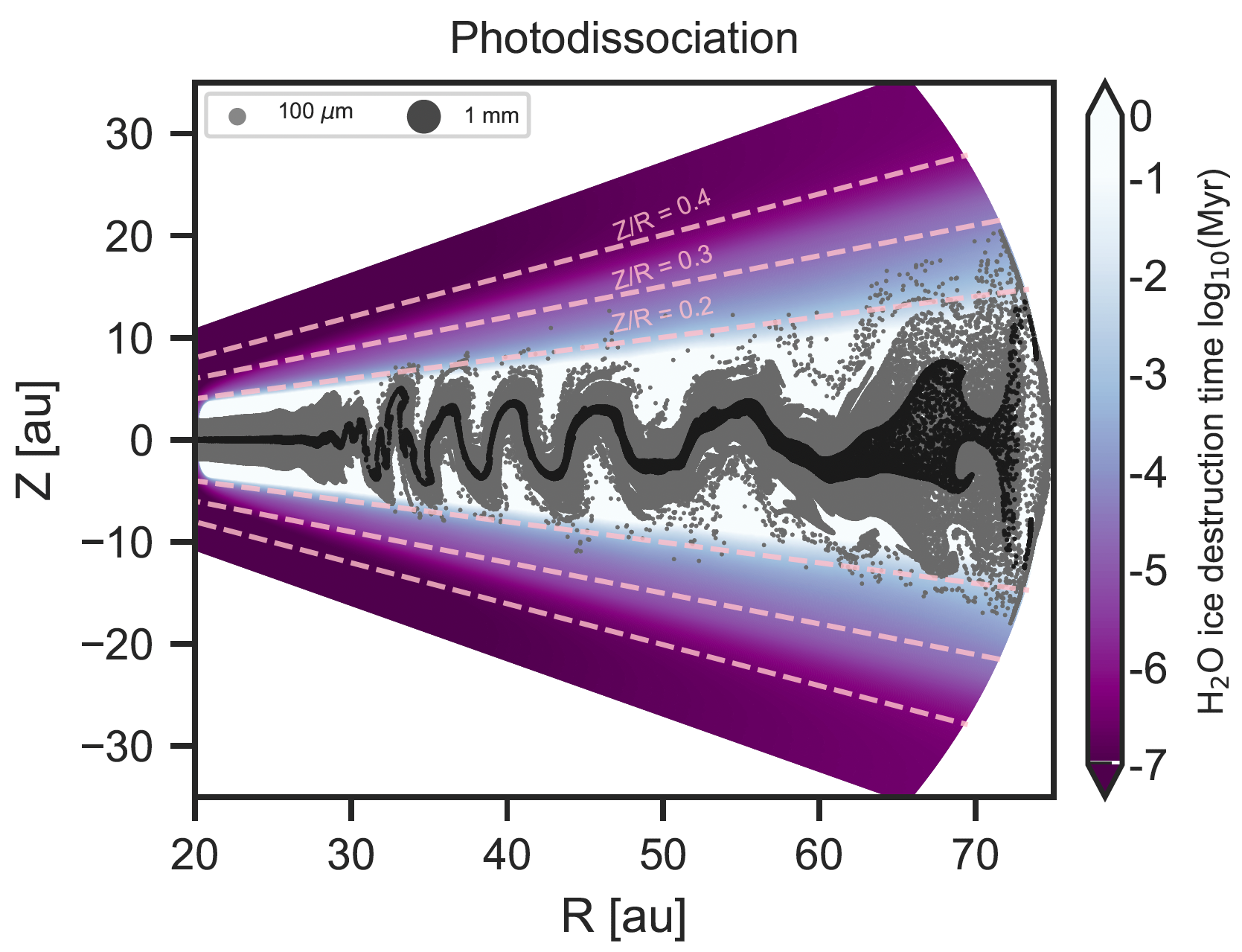}
\caption{H$_{2}$O ice destruction timescale caused by photodissociation in units of Myrs.}
\label{fig:ph_diss_water}
\end{figure}

Figure \ref{fig:ice_destruction_co} (\textit{right}) shows the photodissociation of CO ice in Myrs. The CO ice photodissociation timescale at $Z/R=0.2$ is $t_\mathrm{0.2,phdiss}$$\sim$ 1737.8 years, which is much faster, thus more efficient than photodesorption, to destroy CO ice. For the high stellar luminosity case, all $100~\mu$m particles that cross the $Z/R=0.2$ get photodissociated at a shorter timescale ($t_\mathrm{40,phdiss}$$\sim$ 3.98 years) than $t_\mathrm{40,phdes}$. In the case of H$_{2}$O ice, $t_\mathrm{0.2,phdiss}$ at $Z/R=0.2$ is $\sim 1584.7$ years, which is slightly more efficient than that of CO ice (Figure \ref{fig:ph_diss_water}). It is important to note that both photodesorption and photodissociation reactions are ineffective in the midplane, even in cases of high stellar luminosity. This is because the UV photons already get absorbed by dust particles at $Z/R=0.2$.

As for the small grain sample ($<100~\mu$m), the ice destruction timescale at $Z/R\geq0.2$ is very efficient given that the aerodynamic coupling with the gas will keep the small particles continuously lofted, prolonging settling towards the midplane, and cycled in the upper low-density regions by turbulence. For this reason, ice survival in small particles is expected to be considerably inefficient as they will tend to spend a more significant orbital period than the larger particles in the UV layer.

\subsection{Dynamical timescale of particles at $Z/R=0.2$} 

\begin{figure*}
\centering
\includegraphics[width=9.5cm]{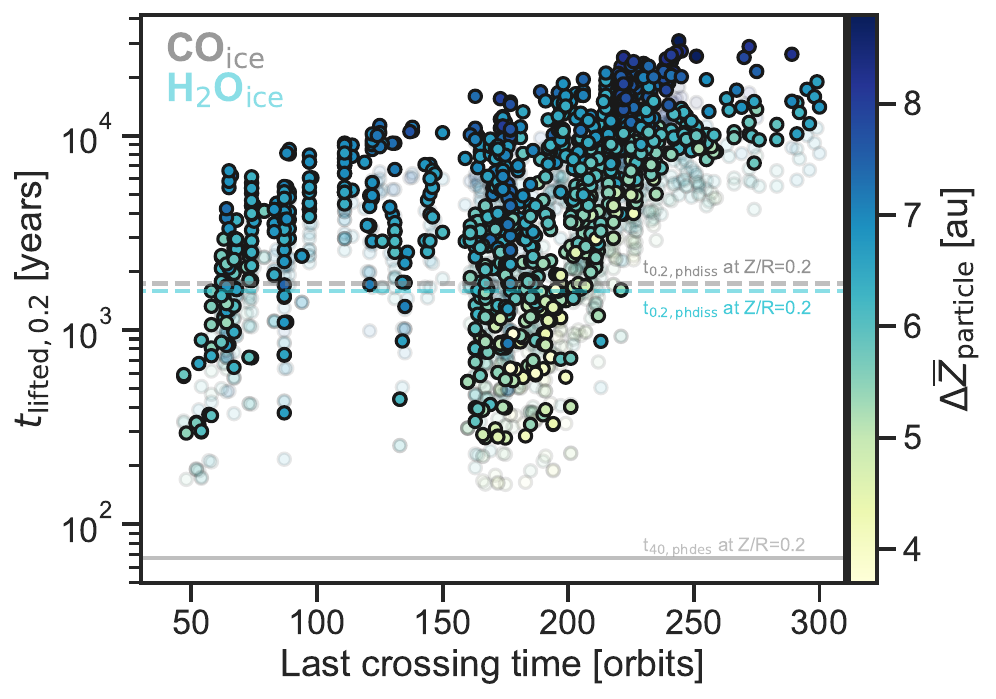}
\includegraphics[width=8cm]{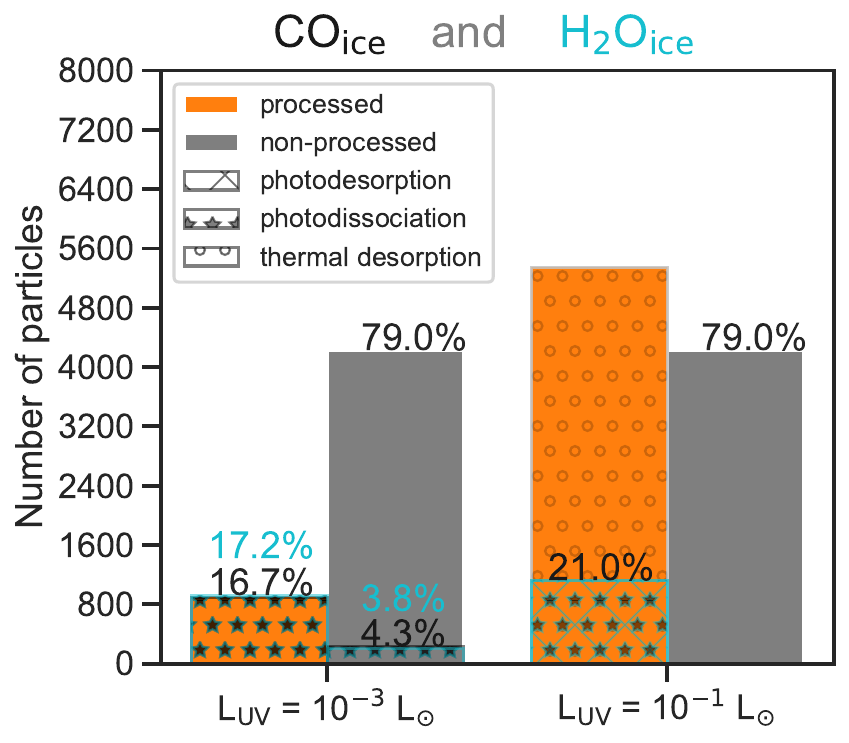}
\caption{\textit{Left.} Particle processing timescales, $t_\mathrm{lifted,0.2}$, compared to the orbital timescales of the $100~\mu$m size particles that cross the $Z/R=0.2$ layer. The x-axis represents the last crossing time in simulation orbit when the particle last crossed the $Z/R=0.2$. The horizontal dashed and solid lines represent the CO ice (\textit{grey}) and H$_{2}$O ice (\textit{light blue}) photoprocesses timescale at $Z/R=0.2$ for the low-luminosity case and the photodesorption timescale at $Z/R=0.2$ for the high-luminosity "test" case, respectively. Particles above these photodestruction timescales are subject to ice processing. The other photodestruction timescales at $Z/R=0.2$ lie outside the $t_\mathrm{lifted,0.2}$ range shown in this plot. The colorbar represents the total distance traveled by the 100~$\mu$m size particles, where the mean total distance peaks around 5.4 au. \textit{Right.} The bar plot shows the distribution of particles processed by thermal or photochemical processes versus those that remain unprocessed despite crossing the $Z/R=0.2$ layer, for both stellar luminosity cases. The sample size comprises 1 mm and $100~\mu$m. In the low-luminosity scenario, 16.6$\%$ and 17.2$\%$ of the crossing particles undergo CO ice and H$_{2}$O processing via photodissociation, all $100~\mu$m particles, respectively. Only 4.3$\%$ and 3.8$\%$ do not experience photoprocessing despite having crossed the $Z/R=0.2$ layer and are added to the non-processed category, which already includes the 79$\%$ UV-shielded particles. For the high-luminosity case, thermal photodesorption affects all particles across all disk radii and vertical layers whereas for photodesorption and photodissociation processes is slightly higher than for the low-luminosity case.}
\label{fig:non_&_processed}
\end{figure*}

In Figure \ref{fig:non_&_processed}, we summarize the processing timescale of all the 1120 particles ($100~\mu$m) that cross the $Z/R=0.2$ layer after the initial dust-settling ($>40$ orbits) to capture the radial and vertical excursion of the particles due to VSI. The \textit{left} plot in Figure \ref{fig:non_&_processed} refers to the ice processing timescale of the particles above the $Z/R=0.2$ layer. The y-axis represents the total orbital time that particles spent above $Z/R=0.2$, and the x-axis represents the last crossing time in simulation orbit when the particle last crossed the $Z/R=0.2$. The last orbital time parameter is helpful to quantify the total number of times the particles have crossed the $Z/R=0.2$ layer (see Figure \ref{fig:factor_time_spent_above_UVlayer}). Note that the shape of the particle distribution follows Figure \ref{fig:factor_time_spent_above_UVlayer}, the occurrence rate of particles crossing $Z/R\geq0.2$. A handful of $100~\mu$m particles have their last $Z/R=0.2$ crossing happening more beyond 150 orbits and before 250 orbits. After 250 orbits, fewer particles have their last crossing at $Z/R=0.2$. This is because settling and inward radial drift occur throughout the entire disk time evolution; thus, the trajectory of the particles will tend to be more settled towards the midplane as the disk continues to evolve in time. We expect only a few to no particles later crossing the $Z/R=0.2$ is expected as after 250 orbits, the dust scale height dynamics reaches equilibrium with the gas. If the number of particles in the statistical analysis is increased, and the simulation is run beyond 300 orbits, the occurrence rate of particles having their last cross at $Z/R=0.2$ can increase, filling in the space after 250 orbits. The colorbar represents the total distance traveled by the 100~$\mu$m size particles, which cover a range from $Z\sim4$~au to $Z\sim8.5$~au. Since some particles cross the midplane, this total distance is calculated based on the average of the positive $Z$s and the negative $Z$s around the midplane for each particle track—the mean total distance peaks around 5.4 au.

The following calculation steps were considered regarding the ice processing timescale of the particles at $Z/R=0.2$, $t_\mathrm{lifted,0.2}$. First, we calculated the orbital period of the 1120 particles at every location when the particles cross the $Z/R=0.2$ layer. This step requires considering the radial distance from the star at every particle radial track point. Because every time the particle crosses the $Z/R=0.2$ has a different radius, we consider the averaged distance traveled by the particle above the $Z/R=0.2$ layer. Second, we recovered the amount of times the particles cross the $Z/R=0.2$ layer (see Figure \ref{fig:factor_time_spent_above_UVlayer} in Appendix).
Similarly to Figure \ref{fig:factor_time_spent_above_UVlayer}, it shows that at most 1120 particles cross the $Z/R=0.2$ layer after 150 orbits of disk evolution time. Finally, we calculated the ice processing time of the particles in years by multiplying the last orbital time of the particle (from the first step) by the amount of times the particles cross the $Z/R=0.2$ layer (from the second step). In other words, if the time the particles spent above the $Z/R=0.2$ layer is greater than the CO ice and H$_{2}$O ice destruction timescale, as calculated in \S\ref{sec:ice_processing}, then the particle is processed. The level of processing of the particles is thus quantified if the particle processing timescales at $Z/R=0.2$, $t_\mathrm{lifted,0.2}$, is greater than any CO ice or H$_{2}$O ice destruction timescale, $t_\mathrm{destroy}$, summarized on the following condition,

\begin{equation}
\centering
\label{processing}
      t_\mathrm{lifted,0.2} \geq t_\mathrm{destroy} = \mathrm{processed} \, 
\end{equation}

It is important to mention that all CO ice and H$_{2}$O ice destruction processes have a different destruction timescale at $Z/R=0.2$. That is why in Figure \ref{fig:non_&_processed} (\textit{left}) we overplotted the CO ice (\textit{grey} line) and H$_{2}$O ice (\textit{blue} line) photodissociation timescale for the low-luminosity case, $t_\mathrm{0.2,phdiss}$, and the photodesorption timescale at $Z/R=0.2$ for the high-luminosity case, $t_\mathrm{40,phdes}$, same for both molecules. The other timescales were not included because 1) the CO ice and H$_{2}$O ice photodesorption timescale for the low-luminosity case is way too long; thus, none of the $100~\mu$m particles get photodesorbed, and 2) the CO ice and H$_{2}$O ice  photodissociation timescales for the high luminosity case is way too fast; less than 4 years all particles have a higher degree of processing at $Z/R\geq0.2$. Given the dependency of the orbital period on stellar mass, we also include the orbital period for the high-mass star shown in Figure \ref{fig:non_&_processed} (\textit{left}) as transparent markers. Note that their orbital timescales determine the ice processing degree in the particle. Therefore, when we say the particle is \textit{processed}, we limit the discussion to whether the particle is \textit{fully} processed or is \textit{partially} processed. It can be intuitive that the faster the photodestruction process is, the more ice will be desorbed from the grain surface. However, this determination will require following the time-dependent chemistry, coupled with other chemical reactions and more species, to accurately determine the amount of CO ice and H$_{2}$O ice that remains on the surface after going through a long track in the disk. This is extremely expensive to do for many particles; however, in \S \ref{sec:time_dependent_chemistry}, we attempted to do this for one particle and generalize the fate for the other particles with the same ice processing timescale history. We expect the particles to be fully processed above the $Z/R=0.4$ layer.

In summary, in Figure \ref{fig:non_&_processed} the bar plot on the \textit{right} shows the number of CO ice (\textit{black}) and H$_{2}$O ice (\textit{light blue}) particles that are processed by any thermal or photochemical processes, and those that, even though they cross the $Z/R=0.2$ layer, do not get processed, for both stellar luminosity cases. The total of particles drawn in the sample was about 5340, which included both 1 mm and 100~$\mu$m size particles. The amount of particles crossing the $Z/R=0.2$ layer is 1120. Thus, the other 4220 particles (mostly 1 mm) remain UV-shielded in the disk. For the low-luminosity case, only 16.7$\%$ and 17.2$\%$ of those 1120, respectively, particles experience some degree of CO ice and H$_{2}$O ice processing caused by photodissociation. We emphasize that all the icy particles processed in the low-luminosity case are 100~$\mu$m size particles. The remaining of 4.3$\%$ and 3.8$\%$ of the 1120 icy particles do not get photoprocessed, which is added up to the non-processed category, already encompassing the 79$\%$ that were already UV-shielded in the disk. No large particles get thermally processed beyond $\sim$30 au for the low-luminosity case. However, inside this radius, all particles get thermally processed for all vertical extensions. On the other hand, for the high-luminosity case, all particles (including those that do not cross the $Z/R=0.2$) get thermally photodesorbed for all disk radius and vertical layers for the same physical disk structure as the low-luminosity case. The photodesorption becomes comparable to the photodissociation where 21$\%$ of the CO ice and H$_{2}$O ice particles crossing the $Z/R=0.2$ layer become processed, while below the $Z/R=0.2$ are the remaining 79$\%$ of particles that are UV-shielded.  

\section{Discussion}

\subsection{Model limitations}

Our hydrodynamical simulations consider the effects of the time evolution turbulence generated during the formation and evolution of a protoplanetary accretion disk. We include the interaction that the gas exerts on the dust particles, but we do not consider the dust-to-gas feedback that significantly impacts the gas flow patterns and can modify local pressure gradients due to dust evolution \citep{Schafer_2022}. Thus, we leave that to future work. The equation of state of the gas dynamics is described in an isothermal configuration, meaning that the local temperature remains constant in the vertical direction. The gas cooling timescale in the outer disk is expected to be much less than one orbital period, making the isothermal configuration a good approximation to describe the gas dynamics in the outer disk. However, these two caveats on the treatment of the hydrodynamics are considered to be less realistic in regions very close to the midplane where the VSI is expected to be less active in the midplane primarily due to stabilizing effects of buoyancy, chemical composition (i.e., high molecular emissivity), dust opacity (i.e., high optical depths trap radiation that reduces the cooling efficiency), and specific thermal conditions (i.e., temperature from the central star) \citep{Fuksman_2024a,Fuksman_2024b}. Our analysis and discussion of the particle dynamics are descriptive of those particles that have not yet coagulated or significantly fragmented with other particles in the disk.

Our ice processing analysis starts by quantifying the radial and vertical excursion of the particles when exposed to time-dependent turbulence in the disk while maintaining computational efficiency good enough to run thousands of analyses of the trajectory of each particle. Our chemical analysis in CO ice and H$_{2}$O ice is in steady-state based on stellar templates for young Class II disks. As the strength level of the UV field in Class 0 and I sources is highly uncertain as well as the vertical height at which UV photons penetrate in the disk, we see that from Figure \ref{fig:physical_structure}, the $Z/R=0.2$ corresponds to the bottom-most layer UV photons could penetrate into. This is also supported by observational evidence of CN emission occupying $Z/R=0.2$ elevations in some disks (more discussion in \S\ref{sec:obs_evi_photochemistry_disks}). During the Class I stage, disks are expected to be puffier, albeit the outflows clear some of it to enable UV-driven chemistry \citep{Flores-Rivera_2021, Flores-Rivera_2023}. However, the disk can still be relatively
embedded, meaning that the UV stellar photons might not reach the $Z/R=0.2$ level as easily as it is during the later
Class II stage. For this reason, we emphasize our ice processing percentages as upper limit values. We do not include the chemical analysis from X-rays and cosmic rays. Regarding the cosmic rays, \citet{Cleeves_2013} found that magnetized winds in T-Tauri stars can significantly reduce the low-energy cosmic ray flux by ten orders of magnitude than the value used for the ISM. Stellar X-ray photons can play an important role by ionizing deeper in the surface layers, particularly when cosmic ray fluxes are reduced, which can increase the ice destruction timescales closer to the midplane.

\subsection{Observational constraints of turbulence and photochemistry in outer disks}
\label{sec:obs_evi_photochemistry_disks}

The vertical extent to which turbulence operates in young disks with ages < 1 Myrs is quite challenging to determine due to the high molecular extinction in highly envelope-embedded disks. The eDisk/ALMA large program has found that some of the sub-mm dusty disks have not yet settled in the midplane \citep{Ohashi_2023}. For example, the IRAS 04302 source has a dust scale height of 6 au at around 100 au disk radius \citep{Lin_2023}. The integrated intensity maps of these young disks show, in general, that Class 0/I disks are more geometrically thick, potentially shaped by strong turbulence. This is the opposite scenario when comparing with Class II disks where the sub-mm dust scale height can be more than a factor of ten lower than the values inferred in Class 0/I disks \citep{Villenave_2020} with turbulence levels as low as $5-10\%$ of the local sound speed \citep{Flaherty_2018, Teague_2018b}, except for IM Lup which shows evidence of strong turbulence at levels of 0.18-0.30 the local sound speed \citep{Flaherty_2024}. However, model predictions are optimistic that young Class II disks ($\sim$ 0.3 Myrs) can experience levels of turbulence of $10^{-3}-10^{-4}$ caused by the VSI and be observationally constrained in the CO upper layers of disks \citep{Barraza-Alfaro_2021}. This prediction still holds when a planet carves a gap in a VSI-active disk. Further high angular (less than 10 au) and spectral ($\sim$0.05 km~s$^{-1}$ with ALMA Band 6) is required to constrain the aforementioned turbulence level \citep{Barraza-Alfaro_2024}. The JWST can also constrain the spatial vertical extent of dust. \citet{Duchene_2024} estimated the thickness of the 890 $\mu$m distribution to be around 14 au (a thickness consistent with our 1 mm and 100 $\mu$m dust excursion heights lifted by the VSI).

The young sources in the eDisk program have systematically analyzed the chemical morphology of some gas molecules to trace various components of the protostellar system \citep[i.e.,][]{Ohashi_2023}. No photochemical imprints have been detected from this sample, therefore, it is uncertain to determine the location of a UV layer. UV photochemistry has been widely studied in Class II T Tauri systems, where the molecular envelope has been mostly dispersed. In situ photochemistry depends on the UV radiation field in which surface layers in the outer disk can still be strongly influenced. However, the vertical extent to which photochemistry is still active is somewhat unclear. The CN/HCN ratio has been proposed to trace UV fields in planetary nebulae and in photodissociation regions \citep{Cox_1992, Fuente_1995}. Motivated by the cross-section dependence of both molecules, recent chemical models have proposed that the most important formation pathway for CN involves the vibrational excited H$_{2}$ that is primarily driven by UV photons, $<105$ nm \citep{Visser_2018, Cazzoletti_2018}. This framework successfully reproduces the CN ring feature detected in TW Hydra \citep{Teague_2016}. Moreover, an excitation analysis of CN emission in TW Hydra \citep{Teague_2017} determined that CN is emitted from low-density, more UV-exposed elevated regions of the disk, consistent with a UV-driven formation pathway. Later, \citet{Bergner_MAPS_2021} conducted a toy model of the local UV field supported by observational evidence and showed that in the region where CN emission is detected, the formation pathway via the vibrational excited H$_{2}$ (91-105 nm) would occupy $Z/R=0.2-0.4$ elevations whereas, beyond $\sim$100 au disk radius, the HCN photodissociation over CN (HCN + $h\nu$ $\rightarrow$ CN + H at 113-150 nm) is more likely to reproduce the observations. Another work by \citet{Benz_2016} explored the role of the UV field based CH$^{+}$/OH$^{+}$ column density ratios to determine that the ratios, analyzed in twelve young sources observed with the \textit{Herschel} Space Telescope, are reproduced for a FUV flux that is 2-400 times greater than that of the interstellar radiation field (ISRF).

The \textit{James Webb} Space Telescope (JWST) has observed ices through spectral absorption features in the envelopes of very young sources. \citet{Brunken_2024} interestingly reported absorption features that appear as two peaks in the splitting bending mode of a couple of Class 0 sources. These absorption features were decomposed into different Gaussian fits augmented with laboratory data to match the spectra. They found that a high fraction of pure $^{12}$CO$_{2}$ ice (at 80 K) in the $^{13}$CO$_{2}$ ice feature indicating ice segregation, where pure $^{12}$CO$_{2}$ separates from other mixed ices due to thermal processing. The concentration of pure $^{12}$CO$_{2}$ ice caused by this process was correlated for sources with high stellar luminosity (as high temperature means a higher segregation process), making the $^{13}$CO$_{2}$ ice a useful tracer of protostellar thermal processing and composition. Furthermore, a work by \citet{Sturm_2023} found that spatially resolved CO ice has been detected at unexpected altitudes within the disk, reaching up to the uppermost layers where CO gas emission is present. The CO ice absorption feature was found to be asymmetric and broader than expected for pure CO ice, which suggests the possible existence of an additional component where CO ice is mixed with either CO$_{2}$ or H$_{2}$O ices, motivating future work to focus on the modeling of a more realistic ice structure (i.e., ice layers) with mixed icy components \citep[see more in][]{Bergner_2024}. In the future, it could be possible for the JWST to detect an ice tracer of photochemistry in young disks. 

\subsection{Comparison with other Lagrangian models in VSI-active disks}

\citet{Flock_2017} used a 3D VSI configuration and treated dust particles in the Epstein regime. In their setup, the Lagrangian dust particles are initially located in the midplane, and as the disk evolves in time, the particles are vertically lifted by the VSI. Our results show that if particles are located initially close to the gas scale height, 0.1 mm particles can be lifted to slightly higher elevations. In our case, the dust scale varies with time, following the oscillations of the VSI, and continues the periodic behavior until the oscillation height reaches equilibrium after $\sim$250 orbits (Figure \ref{fig:time_evol_ke_over_height}) concentrated into a narrow layer of width $\sim$2.3 $H_{R}$ that oscillates around the midplane. In Figure 13 of \citet{Flock_2017}, the mix of the particles is instead more scattered around the midplane, which we attribute to a 3D effect.

Similarly, \citet{Stoll_2016} and \citet{Dullemond_2022} also use a method to treat particles in the Epstein regime by solving the equation of motion of the particles, including the force of gravity and the friction with the gas. They do not follow the tracking of individual particles over time. However, both works showed that the VSI is very effective in isothermal and radiative configurations in stirring up dust particles even with a Stokes number of 0.1 to significant heights equal to the gas pressure scale height, which implies that substantial vertical dust distribution is expected to be seen in observations.

\subsection{Composition of planetary bodies in the comet-forming zone}

The comet-forming zone in the early solar nebula is defined to be in the outer parts of the disk ($>5$ au), beyond the snowline of volatile materials \citep{Whipple_1972, Bockelee_2004}. The smaller grains ($<100~\mu$m) typically undergo significant levels of ice chemical processing above $Z/R>0.2$. As a result, these slow-growing particles experienced a combination of different degrees of ice processing that will eventually be imprinted in the planetary body as isotopic ratio variations \citep{Bergner_2021}. We stress that $100~\mu$m grains are still circulated by turbulence to a lower extent in the UV region, where ice destruction can still play a role by removing some pristine ice from the long-standing orbital period that the dust grains spent above the UV layer. In contrast, larger 1 mm grains are very well shielded in the disk and transported inwards towards the star. The drifting of the particles in the midplane is expected to contain a mix of processed particles inherited from the small grains as they settle toward the midplane. The mm-size grains with a icy mantle composition that have not yet accreted into a planetary-forming body will eventually cross their respective snowline, losing all their pristine composition. 

\citet{van_Kooten_2024} reported the discovery of organic-rich "dark clasts" that represent pebble-sized outer disk materials related to the accretion region of comets. Based on the analysis of their chemical composition, the inferred ice component of this pristine material contains high $\delta^{15}$N and low $\delta$D values compared to pristine ISM ices, which are best explained by vertical mixing of small dust grains ($\leq100~\mu$m) in the outer disk. In detail, these icy grains access the $^{15}$N-rich surface layers of the disk that experience N$_{2}$ photodissociation corresponding to significant N isotope fractionation \citep{Chakraborty_2014}. At the same time, they equilibrate with D-poor H$_{2}$ gas in the disk \citep{Cleeves_2014}, resulting in $^{15}$N-rich and D-poor ice layers coated onto ISM-derived organic matter. This suggests that even the more pristine material from the comet-forming region shows signs of ice processing in its composition. Similarly, small grains are expected to rapidly lose their ice after prolonged exposure in the upper layers. Therefore, it is unlikely that the small particles at these radii are the leading carriers of pristine ISM-like for ice inheritance. Furthermore, they found that the dark clasts have very $^{16}$O-poor bulk compositions that are similar to cometary Interplanetary Dust Particles \citep[IDPs;][]{Starkey_2014, NGUYEN_2022}, suggesting a large addition of outer disk derived ice relative to chondrites, which aligns with the idea of the CO self-shielding isotope effect in the outer disk \citep{Lyons_2005}. We expect that icy bodies, such as comets, inherited mainly their reservoirs of pristine composition from the inward drift and accreting icy pebbles in the midplane. An important future consideration will include dust growth treatment in the simulations to assess the amount of pristine dust aggregates incorporated into the planetary body via streaming instability, SI \citep{Lorek_2018}.

The D/H ratio in water has traditionally served as a tracer for the origin of water on Earth \citep[see review by][]{Ceccarelli_2014}. The D/H ratio of water referenced by the vienna standard mean ocean water (VSMOW) is 1.56$\times$10$^{-4}$ \citep[i.e.,][]{Hallis_2015}, where asteroids have the closest D/H ratio relative to Earth followed by comets \citep{Altwegg_2015}. Variations of the D/H value in comets depend on how much the water is preserved when turbulent transport brings the icy pebbles to the disk surface and photodissociates by UV radiation. Such a photodestruction process in water ice can significantly modify the D/H value in the solar nebula. The influence that water ice photodissociation can have on the D/H ratio would also depend on the presence of other relatively abundant molecules (i.e., atomic oxygen) that help reform water ice back in the disk. \citet{Furuya_2013} found that water ice incorporated into the disk from the molecular cloud is destroyed by UV at the disk surface (at a column density close to $10^{22}$ cm$^{-2}$ around 30 au) while the atomic oxygen entering the deeper disk condenses on the grain surface and converted back to water ice through OH-ice as a reactant, which can increase the water-ice abundance beyond 40 au. In their model with vertical mixing, the water-ice abundance decreases in the midplane as it gets destroyed in the surface layers.
Consequently, water-ice abundance and other deuterated ices (i.e., HDO-ice) decrease over time, and the D/H column density ratio decreases by up to one order of magnitude within $10^{6}$ years between 30 and 50 au. The photoproducts from water-ice, OH-ice, and H-ice can recombine again in the deeper layers with themselves or with H$_{2}$CO-ice to form water ice. The reformation of water ice is efficient as OH-ice should be formed mainly on grain surfaces to compensate for the destruction. In the context of planetesimal composition, the D/H ratio of cometary water has been proven to have originated in the solar nebula given the high D/H ratio detection by JWST \citep{Slavicinska_2024}. However, the D/H ratio decreases due to photoprocessing as the disk evolves. 

The presence of complex organic molecules in protoplanetary disks is crucial for understanding the inventory of prebiotic compounds and their potential survival through the formation and accretion processes that contribute to cometary bodies \citep{Walsh_2014}. Our results indicate that UV photodissociation influences the destruction efficiency of simpler molecules such as CO and water; similarly, we expect the same for complex organics. \citet{Bergner_2021} found that for complex organic molecules rather than HCN, which is the most efficiently photodissociated molecule at Ly-$\alpha$ wavelength, have less Ly-$\alpha$ photodissociation rate suggesting that they are more likely to survive the disk environment.

\subsection{Toy model: surface chemistry evolution for a single particle}
\label{sec:time_dependent_chemistry}

\begin{figure}
\centering 
\includegraphics[width=8.7cm]{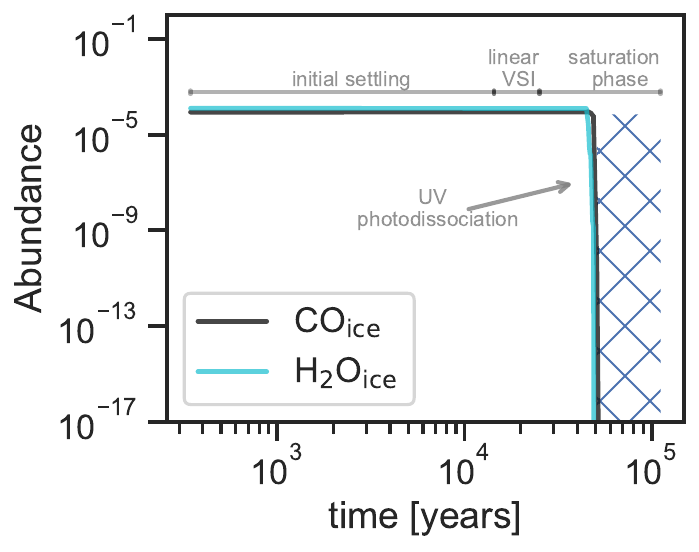}
\caption{Relative abundance of the 100 $\mu$m particle in the first panel in Figure \ref{fig:particle_trajectory_gallery}. The CO$_\mathrm{ice}$ and H$_{2}$O$_\mathrm{ice}$ abundance is a function of orbital time of the particle in years. During the initial dust-settling phase, the CO$_\mathrm{ice}$ and H$_{2}$O$_\mathrm{ice}$ abundance remains constant as the particle remains UV-shielded in the disk until it starts experiencing the vertical circling motion during the VSI non-linear regime at a few tens hundreds of orbits. The big drop in both abundance profiles happen at $\sim50.9$~Kyrs (at 143 orbits at 46.5 au), when the particle reaches the UV-active layer (log$_{10}(\mathrm{F_{UV}})=9.11$~photons s$^{-1}$ cm$^{-2}$) in the saturation-turbulence VSI phase. Here, the particle looses significant amount of CO$_\mathrm{ice}$ and H$_{2}$O$_\mathrm{ice}$ from its surface. These icy volatiles are expected to be completely absent from the grain surface when it is back to the disk midplane ("X" hatched region).}
\label{fig:coice_abundance}
\end{figure}


We follow the surface chemistry of a 100~$\mu$m particle that crosses the $Z/R=0.2$ layer to explore the impact of the particle's trajectory in the disk on the CO$_\mathrm{ice}$ and H$_{2}$O$_\mathrm{ice}$ abundance evolution at the surface of the particle. We use the \textsc{Nautilus} gas-grain astrochemical code \citep{Hersant_2009, Ruaud_2016, Gavino_2021} that considers the rate equation approach based on \citet{Hasegawa_1992}. We used the kinetic database for astrochemistry network \citep[kida.uva.2014;\footnote{\url{https://kida.astrochem-tools.org/networks.html}}][]{Wakelam_2015}. The code includes thermal desorption, cosmic-ray-induced desorption,
photodesorption, and chemical desorption. However, the surface reactions, as well as adsorption, are turned off. There are two reasons to justify this prescription. First, because we want to investigate the impact of the UV field only on the evolution of the CO$_\mathrm{ice}$ and H$_{2}$O$_\mathrm{ice}$, we removed the possible surface reactions that can affect their abundances. Secondly, removing adsorption processes allows us to avoid potential unrealistic re-adsorption events of gas-phase CO and H$_{2}$O when the particles return to regions with lower UV flux. These two prescriptions combined ensure that the CO$_\mathrm{ice}$ and H$_{2}$O$_\mathrm{ice}$ are only affected by the local physical conditions such that we can obtain quantitative information on the UV flux value from which the CO$_\mathrm{ice}$ and H$_{2}$O$_\mathrm{ice}$ photodissociation rate becomes significant.

The \textsc{Nautilus} code considers the physical environment of the particle evolving with time. The code reads the evolving physical parameters (gas density, UV flux, gas temperature, and dust temperature) the particle encountered during its trajectory in the disk. The molecular initial abundances are extracted after an evolution of $10^{6}$ years in a dark molecular cloud (10 K, $n_\mathrm{gas}$ = $10^{4}$, $A_v$=10). The initial abundance used for this molecular cloud simulation is atomic and is the same as those used by \citet{Gavino_2023}.  

Figure \ref{fig:coice_abundance} shows that both the CO$_\mathrm{ice}$ and H$_{2}$O$_\mathrm{ice}$ abundance remain constant as the particle is UV-shielded in the disk during the whole settling and linear-VSI phase. At $\sim46.4$~Kyrs (at 130 orbits at 45.8 au), the particle first enters the UV-active region (log$_{10}(\mathrm{F_{UV}})=7.0$~photons s$^{-1}$ cm$^{-2}$) in the saturation-turbulence phase of the non-linear modes of the VSI. However, the abundances only significantly starts to drop at $\sim50.9$~Kyrs (at 143 orbits at 46.5 au) when the particle reaches a UV value of $9.1$~photons s$^{-1}$ cm$^{-2}$. In agreement with the particle timescale calculation in \S\ref{sec:ice_processing}, the $t_\mathrm{lifted,0.2}=6.1$~Kyrs whereas the UV photodissociation timescale for the $L_{*}$ = 0.2 L$_{\odot}$ case is $t_\mathrm{destroy}=1.3$~Kyrs which means that the particle is subject to ice processing altering the initial composition of the particle.

As mentioned above, some assumptions when reproducing the abundance profiles include the absence of chemical surface reactions. In particular, this excludes the major surface reactions involving CO and water. We performed an additional chemical simulation in which the particle's surface is chemically active, meaning surface reactions without adsorption. We find, for instance, that the activated reaction OH + CO $\rightarrow$ CO$_2$ + H \citep{Ruaud_2015} as well as the reaction CO$_2$ + $h\nu$ $\rightarrow$ CO + O, can significantly change the CO$_\mathrm{ice}$ evolution profile. In the case of H$_{2}$O$_\mathrm{ice}$, it remains adsorbed in grain surface. However, going beyond in the analysis is out of the scope of this study. Adsorption was also excluded in the main simulation (Figure ~\ref{fig:coice_abundance}). To see the impact of adsorption on the evolution of CO$_\mathrm{ice}$, we performed an additional chemical simulation where adsorption is active. In this case, the CO$_\mathrm{ice}$ abundance re-increases when the particle returns to the UV-shielded midplane. But since there is a depletion of the CO gas in the midplane, in principle there should be no CO adsorption in this region ("X" hatched region in Figure \ref{fig:coice_abundance}). This means that a fully processed particle would reach the disk midplane and potentially get mixed with unprocessed icy pebbles.

\section{Conclusions}

Our modeling framework investigates dust particle dynamics and ice processing within a turbulent gas scenario caused by the VSI in the disk. The dust particles are described as Lagrangian particles of 1 mm and 100 $\mu$m in size, experiencing self-consistent gas drag in the disk. The spatial tracks of different particles were examined at different vertical extents. The bottom-most UV layer at $Z/R=0.2$ represents the region where the particles spend enough time to be influenced by UV and thermal ice processing despite being exposed to lower levels of UV radiation, which can still lead to a loss of pristine ice composition in the dust particles. The main points of the conclusion are summarized below.

   \begin{enumerate}
      \item In the scenario where the particles are initially at a certain height above the midplane, the VSI causes significant stirring to them within the disk, with particles of 100 $\mu$m in size frequently crossing the bottom-most UV layer, thereby exposing them to UV radiation that leads to an effective photodissociation. Over time, the percentage of 100 $\mu$ m size particles crossing the UV layer increases, highlighting the impact of VSI on particle dynamics and differential exposure to UV radiation. This suggests that intermediate-sized grains in the upper layers of disks can experience strong replenishment by turbulence in all directions and are expected to be subjected to ice processing due to prolonged exposure to the UV layer. In contrast, larger particles (1 mm) remain shielded below this layer throughout the disk's dynamical evolution, serving as the primary carriers of pristine ice composition within the disk.  

      \item The degree of ice processing in particles within a protoplanetary disk is proportional to the occurrence rate of particles crossing the UV layer. For low-luminosity stars ($L_{*}$ = 0.2 L$_{\odot}$ and $L_\mathrm{UV}$ = 10$^{-3}$ L$_{\odot}$), only 17$\%$ of particles that cross the UV layer experience effective photodissociation. When UV-processed particles return to the midplane, they mix with pristine icy pebbles, leading to a mixed composition.
      
      \item The VSI significantly influences the radial and vertical momentum of dust particles in the disk. As VSI grows over time, it causes particles to experience increasing radial and vertical diffusion. Beyond $\sim$100 orbits, turbulence significantly increases the particle's momentum until it reaches a maximum when in equilibrium with the gas (at $\sim$150 orbits). The resulting dynamical track of the particles is in circular motion, causing the particles to frequently cross the UV layer, prolonging their exposure to UV radiation.      
      
   \end{enumerate}

We highlight how different particle sizes and their initial positions within the disk influence their dynamic behavior and susceptibility to UV-induced ice processing. Our findings suggest that stellar luminosity plays a crucial role in determining the extent of photochemical processing, impacting the evolution and composition of particles within the disk. A future project would follow the chemical analysis for organic and refractory elements. Our findings align with a previous study from \citet{Bergner_2021}, which supports the formation of icy bodies, such as comets, via the accretion of drifting icy pebbles in the outer disk and the inclusion of highly processed material that make up their chemical composition. It further reinforces the role of icy pebbles as the leading volatile carriers, providing critical building blocks to planets where they may contribute to the conditions necessary for the emergence of life.

\begin{acknowledgements}
      We thank the anonymous referee for providing supportive feedback. L. Flores-Rivera and M.L. acknowledge this work is funded by the European Research Council (ERC Starting Grant 101041466-EXODOSS).  S.G. acknowledges support from the Independent Research Fund Denmark (grant No. 0135-00123B). A.J. acknowledges funding from the Danish National Research Foundation (DNRF Chair Grant DNRF159) and the Carlsberg Foundation (Semper Ardens: Advance grant FIRSTATMO). M.F. has received funding from the European Research Council (ERC) under the European Unions Horizon 2020 research and innovation program (grant agreement No. 757957). The Tycho supercomputer hosted at the SCIENCE HPC center at the University of Copenhagen and the astro-node9 cluster at the MPIA were used to support this work. L. Flores-Rivera thanks Troels Haugb\o{}lle for the excellent guidance towards using the Tycho cluster and Sebastian Marino for the useful help regarding some conceptual technicalities when using RADMC-3D. Thanks to Jennifer Bergner and Riohei Nakatani for the helpful discussion about some theoretical chemistry terminology. We thank the matplotlib team \citep{Hunter_2007} for the good quality tool in order to better visualize the data.
\end{acknowledgements}

\appendix

\section{Stellar spectra template comparison}
\label{sec:tw_hydra_spec}

\begin{figure*}
\includegraphics[width=8.9cm]{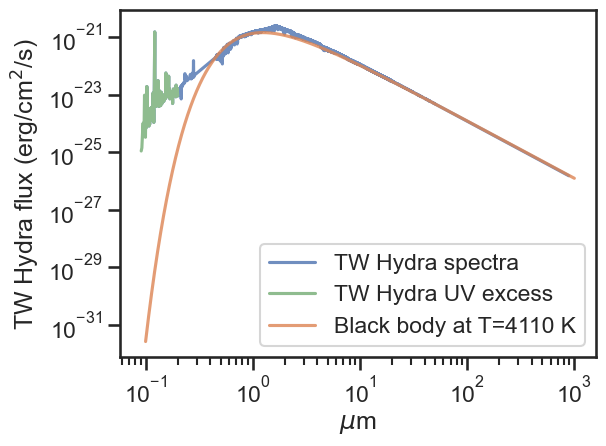}
\includegraphics[width=9cm]{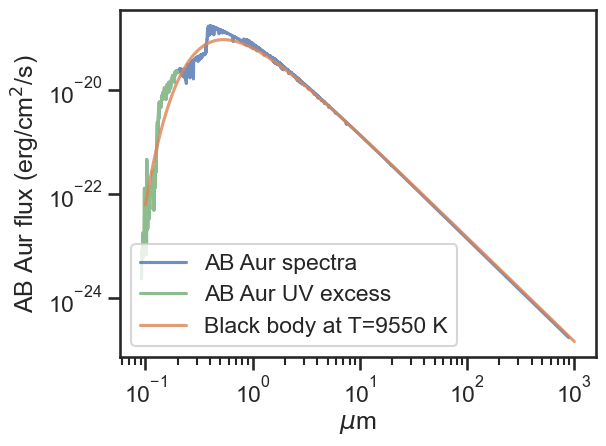}
\caption{TW Hydra and AB Aur stellar spectra templates normalized at 1 pc used as an input parameter in RADMC3D.}
\label{fig:tw_hydra_spec}
\end{figure*}

In addition to TW Hydra, our fiducial sample, we compared it with another case, AB Aur, to explore only the importance of a higher stellar UV field and temperature in the ice processing timescales. Figure \ref{fig:tw_hydra_spec} shows both stellar templates for comparison. The stellar template of AB Aur, a Herbig Ae star, has different stellar properties compared to TW Hydra: $M_{*} = 2.4 ~M_{\odot}$, $T_\mathrm{eff} = 9770$ K, and $R_{*} = 2.37~R_{\odot}$ \citep{Woitke_2016, Boccaletti_2020, Currie_2022}. We recognize that by changing the stellar parameters, the disk properties (i.e., disk temperature and density profiles) must be adjusted; however, doing a parameter study to explore how the chosen stellar properties influence the disk properties is out of the scope of this study. 

\begin{figure}
\includegraphics[width=8cm]{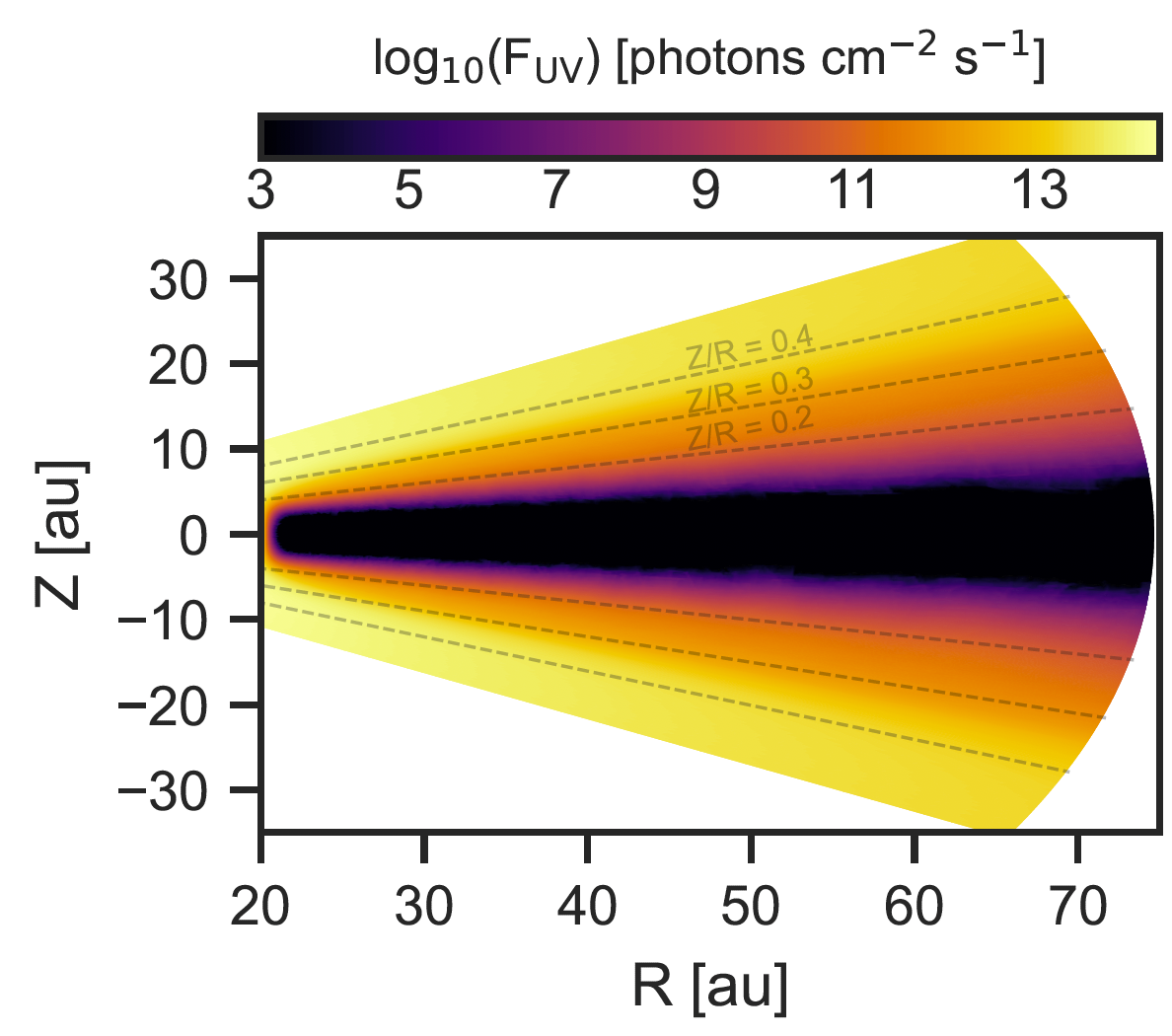}
\caption{AB Aur stellar UV radiation field.}
\label{fig:ab_aur_uv_field}
\end{figure}

Figure \ref{fig:ab_aur_uv_field} shows the UV radiation field of AB Aur, which we use to calculate the CO ice and water ice destruction timescale caused by photodesorption and photodissociation processes. 

\section{Dust opacities}
\label{sec:opacities}

\begin{figure}
\includegraphics[width=8cm]{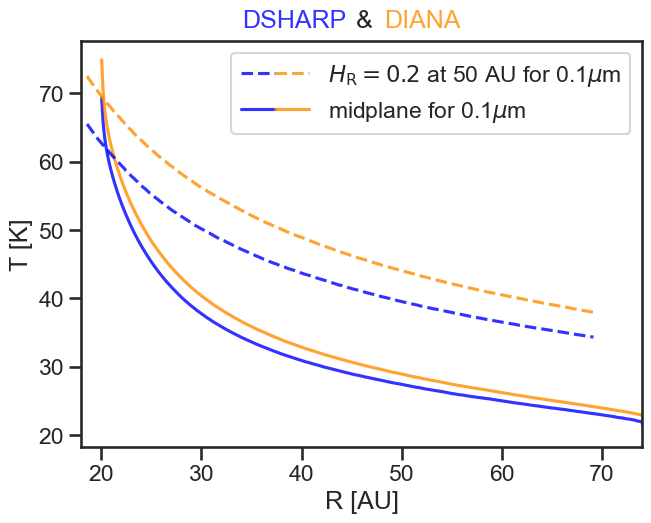}
\caption{Radial temperature profile in the midplane comparison between DSHARP and \textsc{DIANA} opacities.}
\label{fig:opacities}
\end{figure}

We compare the radial temperature profile in the midplane between the DSHARP and the \textsc{DIANA} opacities (Figure \ref{fig:opacities}). The DSHARP opacity follow previous work from \citet{Pollack_1994} regarding protoplanetary disk composition that includes a mixture of water ice ($20\%$ by mass, \citet{Warren_1984}), astronomical silicates \citep{Draine_2003}, Troilite \citep{Henning_1996}, and refractory organics \citep{Henning_1996} without porosity. The \textsc{DIANA} standard dust opacities consider a mixture of $60\%$ amorphous laboratory silicates, $15\%$ amorphous carbon and $25\%$ porosity by volume, well-mixed on small scales \citep{Woitke_2016}.  

\section{Time evolution of the local kinetic energy and dust over height}
\label{sec:time_evol_ke_over_height}

\begin{figure*}
\centering
\includegraphics[width=18cm]{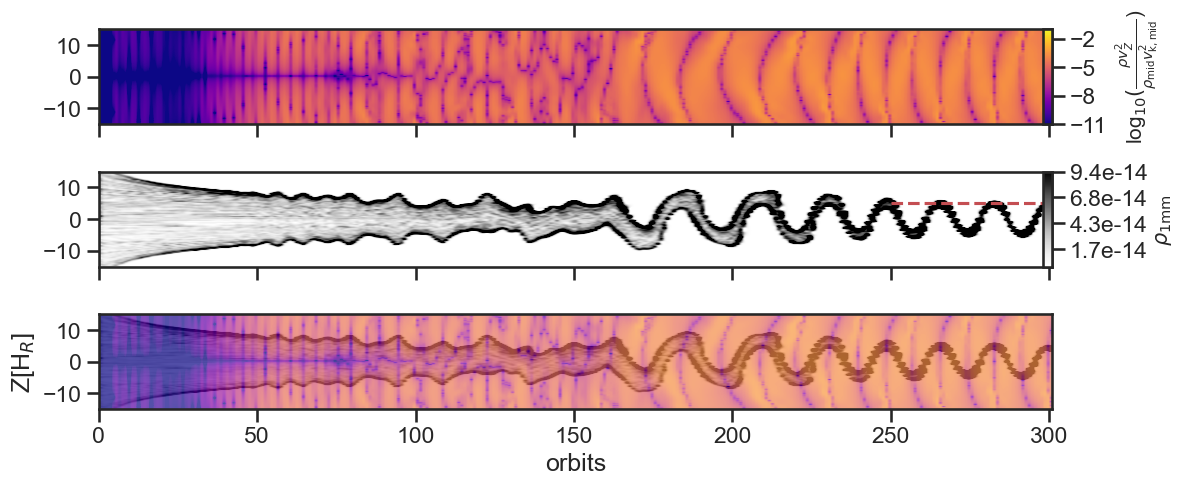}
\caption{Time evolution of the local kinetic energy (\textit{top}) and local 1 mm dust distribution (\textit{middle}) at 50 au. The red-dashed line at Z = 5 H$_\mathrm{R}$ represent the height when particles reach equilibrium with the gas. The \textit{bottom} panel shows the (\textit{middle}) panel overplotted on top of the (\textit{top}) panel.}
\label{fig:time_evol_ke_over_height}
\end{figure*}

We furthermore explore the time evolution of the local kinetic energy normalized with respect to the Keplerian velocity at the midplane, and local dust density at 50 au. In order to map the local kinetic energy and dust density in a more uniform spatial distribution, we transform the velocity vectors to cylindrical and then use a cubic method to interpolate the data on a new grid space with a Z direction ranging from -15 to 15 au at 50 au. As expected, the fast growth of the finger modes in the disk happens within the first few tens of orbits (see Figure \ref{fig:time_evol_ke_over_height} top panel). The dark lane at the midplane shows $v_\mathrm{z}$ is nearly zero. Second energy saturation happens after $\sim$150 orbits. The middle panel in Figure \ref{fig:time_evol_ke_over_height} shows the time evolution of the 1 mm dust density distribution at 50 au. During the second energy saturation, we see particles already going upward and downward and in a circular motion. Notice that during the saturation phase, the particles reach a maximum in height and after 220 orbits y reaches equilibrium with the gas. The bottom panel shows the overlap between the dust density distribution (middle panel) and the local kinetic energy (top panel), both at 50 au.

\section{Occurrence rate of particles above $Z/R=0.2$}
\label{sec:occurence_rate}

\begin{figure}
\includegraphics[width=9cm]{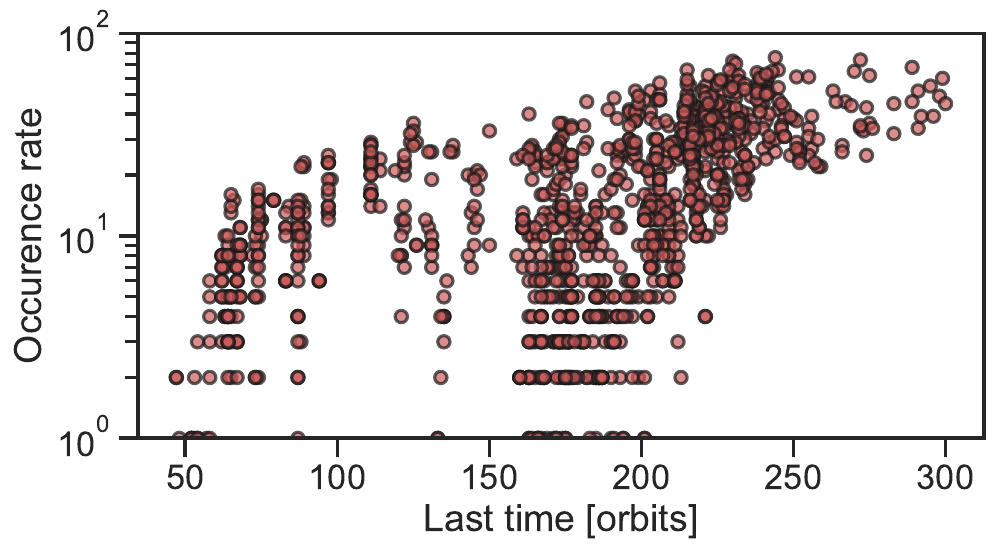}
\caption{Occurrence rate of particles above the $Z/R=0.2$ layer. This occurrence rate represents the amount of times particles spent above $Z/R=0.2$ layer. The x-axis shows time in orbits in which particles made their last UV cross.}
\label{fig:factor_time_spent_above_UVlayer}
\end{figure}

In other to calculate how long do particles spend above the $Z/R=0.2$, we gathered how many times the particles cross this layer. Figure \ref{fig:factor_time_spent_above_UVlayer} shows the amount of times (as a factor) the $100 \mu$m particles spent at $Z/R\geq0.2$.

%
%

\bibliographystyle{aa}
\bibliography{VSI_particles}

\end{document}